\documentclass[lettersize,journal]{IEEEtran}

\usepackage{xcolor}
\colorlet{red}{black}

\usepackage{diagbox}
\usepackage{ulem}

\usepackage{amsfonts}

\usepackage{acronym}

\usepackage{graphicx}

\graphicspath{ {figures/} }

\usepackage{subfigure}

\usepackage[font=footnotesize]{caption}
\usepackage{subcaption}

\usepackage{tikz}

 \usepackage{amsmath, diagbox, hhline, booktabs}
 
 \usepackage[load-configurations=binary,detect-all,binary-units=true,range-phrase=--,per-mode=symbol]{siunitx}
 
 \usepackage[capitalise,noabbrev]{cleveref}

\usepackage{algorithm}
\usepackage{algorithmic}
\newcommand{\tcc}[1]{\textcolor{black}{#1}}
\newcommand{\tcr}[1]{\textcolor{black}{#1}}
\DeclareRobustCommand{\revmark}[1]{#1}

\DeclareRobustCommand{\revdel}[1]{}
\acrodef{6G}{sixth generation}
\acrodef{5G}{fifth generation}
\acrodef{SI}{self-interference}
\acrodef{TX}{transmitter}
\acrodef{RX}{receiver}
\acrodef{ITU-R}{International Telecommunication Union}
\acrodef{IMT-2030}{International Mobile Telecommunication 2030}
\acrodef{ISAC}{integrated sensing and communication}
\acrodef{FD}{full-duplex}
\acrodef{SoI}{signal of interest}
\acrodef{ADC}{analog-to-digital converter}
\acrodef{MIMO}{multiple-input-multiple-output}
\acrodef{CM}{constant modulus}
\acrodef{mmWave}{millimeter wave}
\acrodef{BS}{base station}
\acrodef{MVDR}{minimum variance distortionless response}
\acrodef{RCS}{reflection cross-section}
\acrodef{GD}{gradient descent}
\acrodef{SINR}{signal-to-interference-and-noise ratio}
\acrodef{CM-GD}{constant modulus gradient descent}
\acrodef{PD}{positive definite}
\acrodef{PSD}{positive semidefinite}
\acrodef{OP}{optimization problem}
\acrodef{FP}{fractional programming}
\acrodef{FP-SS}{fractional programming-spherical search}
\acrodef{FP-CSS}{fractional programming-constrained spherical search}
\acrodef{SS} {spherical search}
\acrodef{ES} {exhaustive search}
\acrodef{CSS} {constrained spherical search}
\acrodef{IFP}{integer fractional programming}
\acrodef{IQFP}{integer quadratic fractional programming}
\acrodef{w.r.t.}{with respect to}
\acrodef{CDF}{cumulative distribution function}
\acrodef{SD}{sphere decoding}
\acrodef{RSD}{real sphere decoder}
\acrodef{CSD}{complex sphere decoder}
\acrodef{PSK}{phase shift keying}
\begin{document}
\title{Discrete Codebook Design for Self-interference Suppression in mmWave ISAC}

\author{Guang Chai,~\IEEEmembership{Student Member,~IEEE,} Zhibin Yu,~\IEEEmembership{Member,~IEEE,} Thomas Wagner,~\IEEEmembership{Member,~IEEE,} \\Xiaofeng Wu,~\IEEEmembership{Member,~IEEE,} Giuseppe Caire,~\IEEEmembership{Fellow,~IEEE}
	% <-this % stops a space
	\thanks{This paper was accepted by IEEE Transactions on Wireless Communications on 6 August 2026 [DOI: 10.1109/TWC.2026.3723795]. An earlier version of this paper was presented in part at the 2024 IEEE Global Communications Conference (GLOBECOM) [DOI:10.1109/GLOBECOM52923.2024.10901539].}% <-this % stops a space
	%\thanks{Manuscript received April 19, 2021; revised August 16, 2021.}
	}

%\markboth{Journal of \LaTeX\ Class Files,~Vol.~XX, No.~XX}%
%{Chai \MakeLowercase{\textit{et al.}}: Discrete Codebook Design for Self-interference Suppression in mmWave ISAC}

%\IEEEpubid{0000--0000/00\$00.00~\copyright~2026 IEEE}

\maketitle

\begin{abstract}
This paper presents discrete codebook synthesis methods for \ac{SI} suppression in a mmWave device, designed to support \ac{FD} \ac{ISAC}.
We formulate a \ac{SINR} maximization problem that optimizes the \ac{RX} and \ac{TX} codewords, aimed at suppressing the near-field \ac{SI} signal while maintaining the beamforming gain in the far-field sensing directions. The formulation considers the practical constraints of discrete \ac{RX} and \ac{TX} codebooks with quantized phase settings, as well as a TX beamforming gain requirement in the specified communication direction. 
%To solve the formulated problem, we first fix a properly chosen TX codeword, simplify the problem, and propose a discrete RX codebook design method. The simplified problem is an \ac{IQFP} problem. 
Under an alternating optimization framework, the \ac{RX} and \ac{TX} codewords are iteratively optimized, with one fixed while the other is optimized. When the TX codeword is fixed, we show that the RX codeword optimization problem can be formulated as an \ac{IQFP} problem. Using Dinkelbach’s algorithm, we transform the problem into a sequence of subproblems in which the numerator and the denominator of the objective function are decoupled. These subproblems, subject to discrete constraints, are then efficiently solved by the \ac{SS} method. This overall approach is referred to as FP-SS. 
%Next, we fix a properly chosen RX codeword, simplify the original problem to another IQFP form, and propose a discrete TX codebook design method called the FP-CSS method. 
%While this method also employs the same FP technique, the subproblems are solved using the \ac{CSS} method, which is a further generalization of the previously proposed SS method and incorporates the communication constraint.
When the RX codeword is fixed, the TX codeword optimization problem can similarly be formulated as an IQFP problem, whereas an additional TX beamforming constraint for communication needs to be considered. The problem is solved through Dinkelbach's transformation followed by the \ac{CSS}, and we refer to this approach as FP-CSS. 
We prove that both FP-SS and FP-CSS methods are capable of finding the optimal solutions to their respective TX/RX codebook optimization problems.
Finally, we integrate the FP-SS and FP-CSS methods into a joint TX-RX codebook design approach, where the RX and TX codewords are alternately optimized. 
Simulations show that, 
%the proposed separate RX and TX codebook design methods for the simplified problems can reach the same performance as the corresponding \ac{ES} method, confirming their optimality, but with significantly lower complexity. 
the proposed FP-SS and FP-CSS achieve the same SI suppression performance as their corresponding \ac{ES} methods, confirming their optimality, but at much lower complexity.
%Furthermore, the joint alternating optimization outperforms the individual RX and TX codebook design methods, delivering even better results.
Furthermore, when integrating the proposed FP-SS and FP-CSS methods into the alternating optimization framework, even better SI suppression performance is achieved.       
\end{abstract}

\acresetall
   
   
\acresetall

\section{Introduction}
\label{sec:intro}
In an \ac{ISAC} system, sensing and communication mutually benefit from sharing spectrum and hardware resources~\cite{1}. One of the main \ac{ISAC} use cases is monostatic sensing, where the device operates in the \ac{FD} mode by transmitting a communication signal to the base station while simultaneously receiving the reflection of its own transmitted signals for the purpose of sensing~\cite{2}. Extensive research has focused on monostatic sensing at the infrastructure side (e.g., base stations), where the large antenna size allows sufficient spacing between the \ac{TX} array and the sensing \ac{RX} array, such that \ac{SI} is effectively negligible~\cite{2D,2W}. In contrast, this work focuses on monostatic sensing at the device side, where the compact form factor of handheld or portable devices makes the \ac{SI} non-negligible. 
Due to the limited size of the device and the poor isolation between the \ac{TX} and \ac{RX}, the \ac{SI} signal generated by the transmitter can be orders of magnitude stronger than the received sensing signal. When the power ratio between the received \ac{SI} signal and the sensing signal exceeds the \ac{ADC} dynamic range, digital \ac{SI} cancellation methods cannot be applied. Therefore, sufficient \ac{SI} suppression in the RF analog domain is highly desirable. For \ac{SI} suppression in the analog domain, rather than relying on the design of costly \ac{SI} cancellation circuits~\cite{3}, multi-antenna systems can be explored to suppress SI by beamforming. On one hand, nulls can be introduced by TX beamforming to reduce \ac{SI}~\cite{4,5}; on the other hand, analog phased arrays within a mmWave device can be utilized for \ac{SI} suppression in the analog domain by RX beamforming~\cite{6}, or joint TX and RX beamforming~\cite{8,888,7}.
However, previous research on \ac{SI} suppression through beamforming, as mentioned above, primarily focused on continuous codebook design, without considering the quantization of phase settings. In practice, phase settings are quantized using a limited number of bits, which inevitably leads to performance degradation~\cite{9}. Our recent work in~\cite{0} addresses phase quantization, but it is restricted to RX codebook optimization with fixed TX codewords, which this work extends by jointly optimizing both the TX and RX codebooks.
%In particular, in our previous work~\cite{6,0,888}, instead of only nulling the \ac{SI} signal, we also considered to maintain the beamforming gain at the sensing directions, such that the optimization target becomes the ratio between the sensing power and the \ac{SI} power. The optimization methods in~\cite{6,888} are based on continuous methods wherein the quantization of phase settings is not considered. When the phase settings are quantized with only a few bits, performance degradation cannot be avoided~\cite{9}. The previous work in~\cite{0} addresses phase quantization but is limited to RX codebook optimization with fixed TX codewords, which this work extends by jointly optimizing the TX codebook.
\IEEEpubidadjcol

In this work, as an extension of~\cite{0}, we propose a TX discrete codebook design method and further combine it with the RX discrete codebook design in~\cite{0}, resulting in a joint TX-RX discrete codebook design approach for \ac{SI} suppression. Compared to RX discrete codebook synthesis~\cite{0}, the TX discrete codebook synthesis needs to take into account the communication constraint for TX beamforming. 
The rest of this paper is organized as follows:
As a starting point, we formulate a \ac{SINR} maximization problem, which aims to find the optimal RX and TX codewords, in order to suppress the near-field \ac{SI} while maintaining the beamforming gain at the far-field sensing directions. The optimization problem is constrained by the \ac{CM} and discrete phase requirements of the device’s RX and TX phased arrays, as well as a communication constraint that the TX beamforming gain at the given communication direction should be larger than or equal to a desired value. 

Then, we solve the formulated problem in steps. First, we focus on the optimization of the discrete RX codebook while keeping the TX antenna weight vector fixed. This leads to a simplified problem, which is an \ac{IQFP} problem. To solve this simplified problem, we apply the \ac{FP} technique, Dinkelbach's algorithm~\cite{15a,15}, to transform the problem into a sequence of subproblems where the numerator and the denominator of the objective function are decoupled. When assuming uniformly quantized phase settings, the discrete CM codewords lie on a \ac{PSK} constellation. With this intuition, we show that the subproblems, related to discrete phase finding for RX codebook design, are similar to the PSK data symbol demodulation problem. Hence, the subproblems are then solved by a low-complexity spherical search (SS) method, which is inspired by the \ac{CSD}~\cite{a10,aa10} originally designed for low-complexity PSK data demodulation. This overall approach combining the FP technique and the \ac{SS} method is referred to as the FP-SS method. We prove that the proposed FP-SS method is able to find the optimal solution to the simplified problem, and that Dinkelbach's algorithm converges superlinearly in the FP-SS method.

Next, we focus on the optimization of the discrete TX codebook for a fixed RX antenna weight vector. The resulting simplified problem again has an IQFP form which can be transformed into a sequence of subproblems using the FP technique. %To solve the simplified problem the same \ac{FP} technique is applied, but the subproblems are solved using the \ac{CSS} method, which is a generalization of the \ac{SS} method and additionally takes the communication constraint into account. 
Compared to the RX discrete codebook synthesis problem, where the subproblems can be solved by the SS method, the TX discrete codebook synthesis additionally needs to consider a TX communication constraint. To handle this constraint, we extend the SS method and propose a generalized version, referred to as the CSS method.
The overall approach is named as FP-CSS. 
%It can be proved, in the same manner as the FP-SS method, that the FP-CSS method is capable of finding the optimal solution to the corresponding simplified problem.
%In particular, to enable the \ac{SS} and \ac{CSS} methods, we explore the \ac{CM} property of the codewords to generate a \ac{PD} matrix structure, such that the Cholesky decomposition~\cite{aa9} is guaranteed to be possible~\cite{a9}. 

Finally, we propose the joint optimization of both TX and RX codebooks by alternating between the FP-SS and FP-CSS methods, where the RX and TX codewords are iteratively optimized. 

Simulations are conducted with realistic antenna array placements and phase quantization for a mmWave ISAC device. The results show that, the proposed FP-SS and FP-CSS methods can reach the same performance as their corresponding \ac{ES} methods, which aligns with the proved optimality, but at much lower complexity. Furthermore, the joint codebook design method outperforms the individual RX and TX codebook design methods, yielding even better SI suppression performance. 

Notation: Bold lowercase $\mathbf{a}$ is used to denote column vectors, bold
uppercase $\mathbf{A}$ is used to denote matrices, non-bold letters $a$, $A$
are used to denote scalar values. Using this notation, $|a|$ is the magnitude of a scalar, $\left \| \mathbf{a} \right \|_2$ is the $2$-norm of vector $\mathbf{a}$, $\mathbf{A}^{H}$ is the conjugate transpose, $\mathbf{A}^{T}$ is the matrix transpose, $\mathbf{I}_n$ denotes the identity matrix of dimension $n\times n$, $[\mathbf{A}]_{n,m}$ is the element in the $n$-th row and $m$-th column of $\mathbf{A}$. $arg(\cdot)$ returns the phase of a complex number. $\text{vec}(\cdot)$ denotes the vectorization operation that converts the matrix into a vector in the order of column, and $\text{mat}(\cdot)$ is the inverse operation of $\text{vec}(\cdot)$ that converts the vector into a matrix.
\begin{figure}[t]
	\centering
	\includegraphics[trim={1cm 0cm 0.8cm 0.8cm},clip, width=0.47\textwidth]{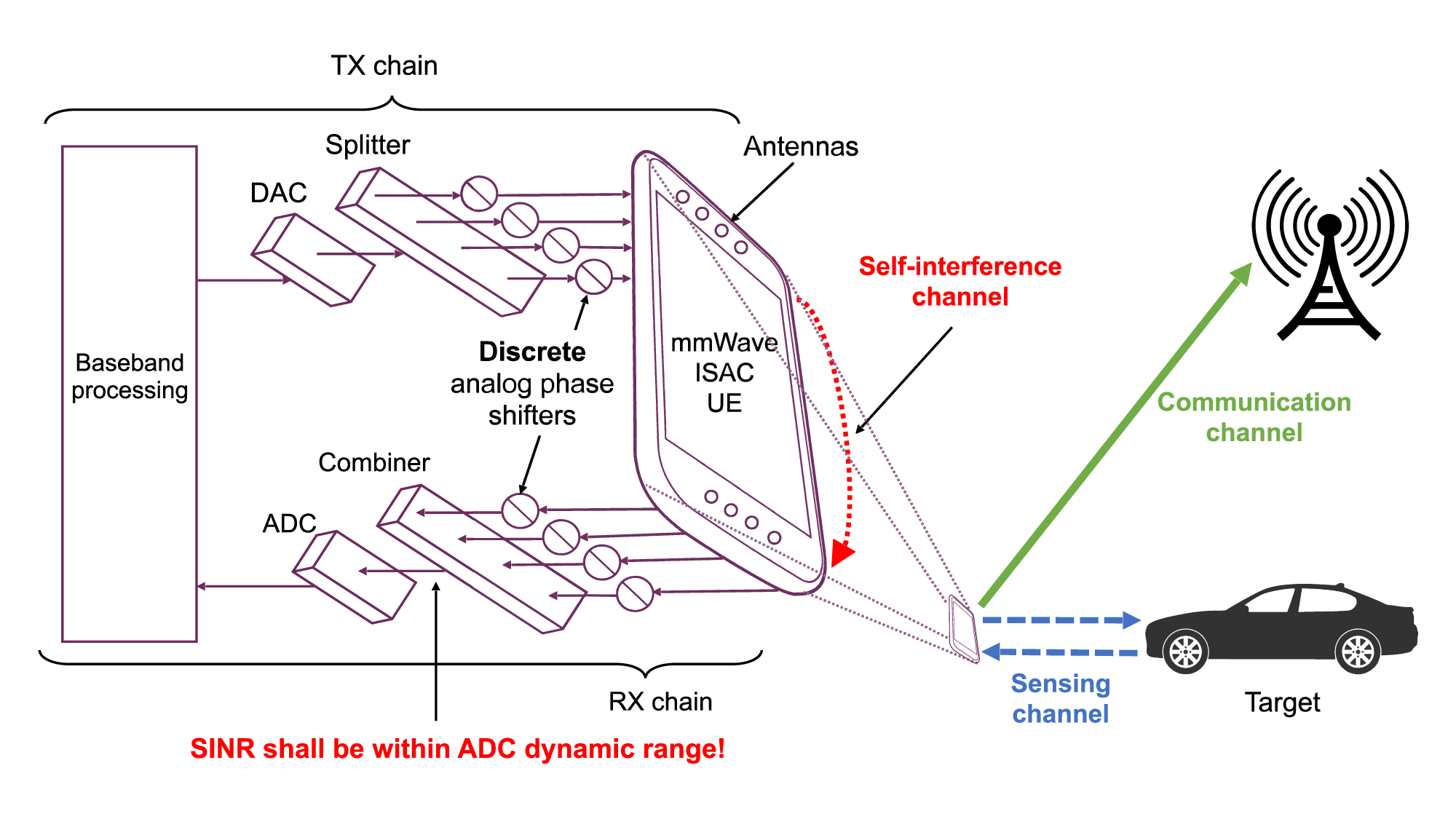}
	%\vspace{-5mm}
	\caption{An ISAC terminal device equipped with a mmWave MIMO transceiver with discrete analog phase shifters.}
	%(Figure is not to scale and only relevant components are shown for clarity).}
\label{fig:int1}
%\vspace{-8mm}
\end{figure}
\section{Problem Formulation}
\label{sec:system}

In this section, we first present a system model where the \ac{SI} signal propagates through a near-field channel, while the sensing and communication signals propagate through a far-field channel. We assume the channel bandwidth $W$ is sufficiently small, i.e., $W\ll$ $f_0$ (the carrier frequency). Thus, we can approximate the carrier wavelength $\lambda$ as $\lambda=\frac{c}{f_0}$ across the entire bandwidth, where $c$ is the speed of light. Subsequently, we formulate a \ac{SINR} maximization problem, which is constrained by discrete phase settings and a communication constraint.

\subsection{System Model}
\label{subsec:channel}
%\vspace{-2mm}

We consider a \ac{mmWave} ISAC device equipped with a TX array of $M$ antennas and an RX array of $N$ antennas, each associated with a single RF chain. The device operates in \ac{FD} mode, simultaneously transmitting a communication signal to the base station and receiving the backscattered signal for sensing.
The received signal $y$ prior to entering the \ac{ADC} is\footnote{\tcr{In practical FD or ISAC transceivers, the SI signal, which could easily cause ADC saturation due to its strong power, contains both linear and nonlinear components. Typically, the linear SI component is much stronger than nonlinear terms~\cite{EE}. Therefore, analog cancellation is first applied to suppress the linear SI and prevent ADC saturation, while digital cancellation is subsequently used to further suppress the residual SI, including nonlinear terms~\cite{AA,AAB}. For these reasons, this work for analog cancellation adopts a linear SI model, which is also widely used in prior mmWave FD and ISAC studies~\cite{BB,CC,DD}.}}:
\begin{align}\label{eq:y}
y=\mathbf{w}^{H}\mathbf{H}_{\mathrm{sens}}\mathbf{v} x+\mathbf{w}^{H}\mathbf{H}_{\mathrm{SI}}\mathbf{v} x+\mathbf{w}^{H} \mathbf{z},
\end{align}
where $x$ represents the transmitted signal, $\mathbf{w}=[e^{j \varphi_1 },...,e^{j \varphi_{N-1}},1 ] ^T\in \mathbb{C}^{N\times 1}$ and $\mathbf{v}=[e^{j \psi_1 },...,e^{j \psi_{M-1}},1 ] ^T\in \mathbb{C}^{M\times 1}$ are the \ac{CM} RX and TX beamforming vectors, respectively. Considering discrete phase settings, we define the set of possible phases $\mathcal{D}=\{0, \Delta ,...,2\pi-\Delta \}$ with uniform quantization, where $\Delta=\frac{2\pi}{2^b}$, and $b$ is the quantization bit number. Thus, $\varphi_n, \psi_m \in \mathcal{D}$ for $n=1, 2,  ..., N$ and $m=1, 2,  ..., M$. Since only the phase difference is relevant in \ac{CM} beamforming, $\varphi_N$ and $\psi_M$ are set to $0$. Hereafter, we refer to a \ac{CM} beamforming vector as a codeword.
$\mathbf{H}_{\mathrm{sens}}\in \mathbb{C}^{N\times M}$ and $\mathbf{H}_{\mathrm{SI}}\in \mathbb{C}^{N\times M}$ represent the sensing and \ac{SI} channel matrices, respectively. Moreover, $\mathbf{z}\in \mathbb{C}^{N\times 1}$ is a zero-mean complex Gaussian noise vector with covariance matrix $\sigma_z^{2}\mathbf{I}_N$, i.e., $\mathbf{z} \sim \mathcal{CN}\left(\mathbf{0}, \sigma_z^{2}\mathbf{I}_N\right)$. $\mathbf{H}_{\mathrm{sens}}$ is modeled under the far-field assumption as~\cite{a11,aa11}: 
\begin{align} \label{eq:3}
\mathbf{H}_{\mathrm{sens}}=  \sum_{l=1}^{L} \alpha _{l} \mathbf{a}_r\left(\theta_{l}\right)\mathbf{a}_t^{H}\left(\phi_{l}\right),
\end{align}
where $L$ is the total number of single-bounce reflection paths, including reflections from both the targets of interest and clutters in the environment. Multi-bounce reflection paths are assumed to be too weak to consider. 
Each path $l$ is characterized by the angle of arrival $\theta_{l}$, the angle of departure $\phi_{l}$, and the complex total path gain $\alpha _{l}$. Without loss of generality, we assume a uniform linear array (ULA) configuration for both the TX and RX, with steering vectors represented by $\mathbf{a}_t(\phi_{l})$ and $\mathbf{a}_r(\theta_{l})$, respectively.
For an $N$-element ULA with half-wavelength spacing, the steering vector\footnote{In Section~\ref{sec:robustness}, we explicitly assess the sensitivity of the sensing SINR to array manifold error (AME) by applying independent per-element gain and phase perturbations~\cite{JJ,KK} to the far-field sensing steering responses.} $\mathbf{a}(\theta)\in \mathbb{C}^{N\times 1}$ is given by:
\begin{align} 
\mathbf{a}(\theta)=\left[1, e^{j \pi \sin(\theta)}, e^{j \pi 2\sin(\theta)}, \ldots, e^{j \pi (N-1)\sin(\theta)}\right]^{T}.
\end{align} 
We assume that the TX and RX ULAs are placed parallel to each other and operate in a monostatic mode, hence $\phi_{l}=\theta_{l}$.

The amplitude of the complex total path gain $|\alpha_{l}|$ is~\cite{11}:
\begin{align} 
|\alpha_{l}| =   \sqrt{\frac{ G_{t,\theta _{l}}G_{r,\theta _{l}} \lambda^{2}}{(4 \pi)^{3} r_{l}^{4}} \sigma_{l} },
\end{align}
where $\lambda$ is the wavelength, 
$r_{l}$ is the distance to the  scatterer on the $l$-th path,  $\sigma_{l} $ is the \ac{RCS} of scatterer $l$, and $G_{t,\theta _{l}}$ and $G_{r,\theta _{l}}$ are the TX and RX antenna gain at $\theta _{l}$ for each single antenna element, respectively.

\tcc{For compact user-end ISAC devices, the TX and RX antennas are separated by only several centimeters, and the SI is dominated by the near-field coupling between TX and RX antennas. 
Therefore, we model the \ac{SI} channel by modeling this near-field coupling.} 
%The extension by considering the NLOS paths caused by clutters can be considered for future work. 
The \ac{SI} channel coefficient between the $n$-th RX antenna and $m$-th TX antenna $[\mathbf{H}_{\mathrm{SI}} ]_{n,m}$, with the distance $d_{n,m}$ between them, is modeled as~\cite{12,13}:
%[\mathbf{H}_{\mathrm{SI}} ]_{n,m}$ represents the SI channel coefficient between the $n$-th RX antenna and the $m$-th TX antenna, and it is given by~\cite{12,13}:
\begin{align}
[\mathbf{H}_{\mathrm{SI}} ]_{n,m} =\sqrt{\beta_{n,m}} \cdot  e^{-j\frac{2\pi}{\lambda } d_{n,m} }.
\label{eq:HSI_nm}
\end{align}
Here, $\beta_{n,m}$ is the underlying coupling coefficient when the antennas share the same substrate, and it is given by~\cite{14}:
\begin{align}
\beta_{n,m}=\frac{G_{3}^{2} \lambda^{2}}{16 \pi^{2} d_{n,m}^{2}}+\frac{G_{2}^{2} \lambda}{4 \pi^{2} d_{n,m}}+\frac{2 G_{3} G_{2} \lambda^{3 / 2}}{8 \pi^{2} d_{n,m}^{3 / 2}},
\label{eq:beta_nm}
\end{align}
where $G_{2}$  is the two-dimensional antenna gain corresponding to surface propagation and $ G_{3}$ is the three-dimensional antenna gain corresponding to free-space propagation.

\noindent\textcolor{red}{\textbf{Remark:}
Although the nominal SI channel in this work is generated from the near-field coupling model in Equations~\eqref{eq:HSI_nm}--\eqref{eq:beta_nm}, the proposed FP-SS and FP-CSS algorithms are not tied to this specific parametric model. Whether the SI channel matrix \(\mathbf{H}_{\mathrm{SI}}\) is obtained from the near-field coupling model, offline calibration, or online channel estimation, the same FP-SS/FP-CSS procedures can be applied without modification. Section~\ref{sec:robustness} provides an example with composite SI channels including reflected NLoS components.}

\subsection{\ac{SINR} Maximization}
\label{subsec:SINR Maximization}
In~(\ref{eq:y}), the received sensing signal and the SI signal are correlated due to the transmitted signal $x$. Fortunately, since the SI channel has a shorter delay, the two signals can be separated in the delay domain. However, if the power ratio between the two signals exceeds the ADC's dynamic range, digital processing required for delay-domain separation becomes infeasible. Furthermore, when the fractional delay of the SI channel lies within the ADC's sampling intervals, the strong power of the SI signal can spread over the delay domain, making the separation extremely difficult. For these reasons, it is highly desirable to suppress the SI signal power while maintaining the sensing signal power prior to ADC sampling. Motivated by this, we formulate an SINR maximization problem.

Based on the sensing channel model from Equation~(\ref{eq:3}), the received sensing signal can be written as:
\begin{align}
\mathbf{w}^{H}\mathbf{H}_{\mathrm{sens}}\mathbf{v} x=x\sum_{l=1}^{L}\alpha _{l} \mathbf{w}^{H}\mathbf{a}_r\left(\theta_{l}\right)\mathbf{a}_t^{H}\left(\theta_{l}\right)\mathbf{v},
\end{align} 
where $\alpha _{l} \mathbf{w}^{H}\mathbf{a}_r\left(\theta_{l}\right)\mathbf{a}_t^{H}\left(\theta_{l}\right)\mathbf{v}$ characterizes the beam amplitude of the reflection from the scatterer $l$, 
while $\mathbf{w}^{H} \mathbf{H}_{\mathrm{SI}}\mathbf{v}$ in~(\ref{eq:y}) characterizes the \ac{SI} amplitude after beamforming.
We define $\theta$ as a sensing direction to explore the environment. Since the device is unaware of the actual directions of the targets, it performs beam sweeping by forming RX/TX beam pairs at various potential sensing directions. Each beam pair formed for a given sensing direction should simultaneously achieve SI suppression, provide sufficient beamforming gain in the respective sensing direction, and maintain adequate beamforming gain for communication. For the codebook design, since the path gain of the actual reflected path at $\theta$ is unknown, we assume a worst-case path gain $\alpha_{\theta}$, which corresponds to the minimal reflected power from the tentative target at the sensing direction $\theta$. $\alpha_{\theta}$ can be calculated based on the sensing requirements of the \ac{ISAC} terminal device, such as the worst-case sensing distance and the worst-case \ac{RCS} of the tentative target at $\theta$. With this assumption, the \ac{SINR} at the sensing direction $\theta$ can be defined as follows:  
\begin{equation}\label{eq1}
\begin{split}
	\text{SINR}_{\theta}(\mathbf{w},\mathbf{v})
	&=\frac{P_{t}\left |\alpha_{\theta} \mathbf{w}^{H}\mathbf{a}_r\left(\theta\right)\mathbf{a}_t^{H}\left(\theta\right)\mathbf{v} \right |^{2}}{P_{t}\left | \mathbf{w}^{H} \mathbf{H}_{\mathrm{SI}}\mathbf{v} \right |^{2}+ N\sigma_z^{2}}
\end{split}
\end{equation} 
where $P_{t}$ is the transmission power at each TX antenna. The task of codebook synthesis can be defined as maximizing $\text{SINR}_{\theta}(\mathbf{w},\mathbf{v})$. We consider the communication constraint that the TX beamforming gain at a given communication direction $\theta_c$ must be greater than or equal to a specified value $c$. The \ac{SINR} maximization problem at $\theta$, \ac{w.r.t.} the \ac{CM} beamforming vectors $\mathbf{w}$ and $\mathbf{v}$, is then formulated as: 
%\vspace{-0.5mm}
\begin{equation}\label{op}
	\begin{split}
		\underset{\varphi_1,...,\varphi_{N-1},\psi_1,...,\psi_{M-1}}{\max} \quad  \frac{\left | \mathbf{w}^{H}\mathbf{a}_r\left(\theta\right)\mathbf{a}_t^{H}\left(\theta\right)\mathbf{v} \right |^{2}}{\left | \mathbf{w}^{H} \mathbf{H}_{\mathrm{SI}}\mathbf{v} \right |^{2}+ \frac{N}{P_{t}}\sigma_z^{2}}, 
		\\
		\text{s.t.}\ \  |\mathbf{v}^{H}\mathbf{a}_{t}(\theta_{c})|^2 \geq c^2,
	\end{split}
\end{equation}
where $\left |\alpha_{\theta}\right |^2$ is omitted from the objective function since it
does not affect the solution to the optimization problem. We denote the objective function in \ac{OP}~(\ref{op}) as $q_{\theta}(\mathbf{w},\mathbf{v})$. 
%\sout{The task of codebook synthesis can be defined as}
%\sout{maximizing $\text{SINR}_{\theta}(\mathbf{v},\mathbf{w})$.}

In~(\ref{op}), we assume that the value of \( c^2 \) is chosen such that~(\ref{op}) remains feasible. Specifically, \( c^2 \) should lie within the range \([0, M^2]\), where the upper bound \( M^2 \) is generally not attainable for arbitrary \( \theta_c \) and quantized phase values of \( \mathbf{v} \).  
To address this, we select \( c^2 \) as a fraction of the maximum theoretical beamforming gain \( M^2 \). Notably, \( c^2 \) serves as a tuning parameter that balances the trade-off between communication and sensing performance.  
For instance, choosing \( \mathbf{v} \) to maximize \( \left| \mathbf{v}^H \mathbf{a}_t(\theta_c) \right|^2 \) and letting \( c^2 \) equal to this maximum yields that \( \mathbf{v} \) is fixed and the objective function in~(\ref{op})  can only be optimized \ac{w.r.t.} \( \mathbf{w} \).   
On the other hand, if \( c^2 \) is set to a value lower than \( \max_{\mathbf{v}} \left| \mathbf{v}^H \mathbf{a}_t(\theta_c) \right|^2 \), then~(\ref{op}) can be effectively optimized \ac{w.r.t.} both \( \mathbf{v} \) and \( \mathbf{w} \), making the communication constraint non-trivial.

\section{Discrete RX Codebook Design with a Fixed TX Codeword}\label{RX}

In this section, we assume that the TX codeword is fixed, which leads to a simplified problem where only the RX codeword needs to be optimized. We show that this simplified problem can be formulated as an \ac{IQFP} problem. We then apply the FP technique, Dinkelbach's algorithm, to transform the problem into a sequence of subproblems, where the numerator and the denominator of the optimization target are decoupled. Next, we show that these subproblems can be efficiently solved by the \ac{SS} method. Finally, we prove that the proposed FP-SS method, which combines the \ac{FP} technique (specifically Dinkelbach's algorithm) and the \ac{SS} method, can find the globally optimal solution of the simplified problem and is guaranteed to converge.

 We assume that the TX codeword $\mathbf{v}$ is fixed as $\mathbf{v}=\mathbf{v}_{0}$, which fulfills the constraint of discrete phase settings and the communication requirement. Hence, the task is to find the optimal RX codeword $\mathbf{w}$ that can maximize the receiver SINR given that $\mathbf{v}=\mathbf{v}_{0}$. 
Denote $\mathbf{b}(\theta)=\mathbf{a}_r(\theta)\mathbf{a}_t^{H}(\theta)\mathbf{v}_{0}\in \mathbb{C}^{N\times 1}$ and $\mathbf{g}_{\mathrm{SI}}=\mathbf{H}_{\mathrm{SI}}\mathbf{v}_{0}\in \mathbb{C}^{N\times 1}$. 
%\sout{Note that $\mathbf{b}(\theta)$ is $\mathbf{a}_{r}(\theta)$ scaled by a complex number.}
The objective function in~(\ref{op}), $q_{\theta}(\mathbf{w},\mathbf{v}=\mathbf{v}_{0})$, is reduced to: 
\begin{align}q_{\theta}'(\mathbf{w})
	&=\frac{\left | \mathbf{w}^{H}\mathbf{b}(\theta) \right |^{2}}{\left | \mathbf{w}^{H} \mathbf{g}_{\mathrm{SI}} \right |^{2}+ \frac{N}{P_{t}}\sigma_z^{2}}\\
	&=\frac{\left | \mathbf{w}^{H} \mathbf{b}(\theta) \right |^{2}}{\mathbf{w}^{H}(\mathbf{G}_{\mathrm{SI}}+\frac{\sigma_z^{2}}{P_{t}}\mathbf{I}_N)\mathbf{w}},
\end{align} 
where $\mathbf{G}_{\mathrm{SI}}=\mathbf{g}_{\mathrm{SI}}\mathbf{g}_{\mathrm{SI}}^{H}\in \mathbb{C}^{N\times N}$. We further define $\mathbf{G}=\mathbf{G}_{\mathrm{SI}}+\frac{\sigma_z^{2}}{P_{t}}\mathbf{I}_N\in \mathbb{C}^{N\times N}$ and $\mathbf{B}(\theta)= \mathbf{b}(\theta)\mathbf{b}(\theta)^{H}\in \mathbb{C}^{N\times N}$. Note that $\mathbf{B}(\theta)$ is \ac{PSD} and $\mathbf{G}$ is \ac{PD}. The \ac{SINR} maximization problem at the sensing direction $\theta$ \ac{w.r.t.} the codeword $\mathbf{w}$ is:
%\vspace{-0.5mm}
\begin{align}\label{eq:obj2}
	\underset{\varphi_1,...,\varphi_{N-1} }{\max} \frac{\mathbf{w}^{H}\mathbf{B}(\theta)\mathbf{w}}{\mathbf{w}^{H}\mathbf{G}\mathbf{w}}, \  \text { s.t. } \varphi_1,...,\varphi_{N-1} \in \mathcal{D},
\end{align}
 which is an \ac{IQFP} problem. In the following, we present the proposed FP-SS method for solving the problem~(\ref{eq:obj2}). First, the FP method is applied to transform the problem into a sequence of subproblems, where the numerator and the denominator of the optimization target are decoupled. Then, we show that the subproblems can be efficiently solved by the \ac{SS} method. 
\subsection{SINR Maximization by the FP Method}
\label{subsec:refinement} 
A single-ratio maximizing \ac{FP} problem is defined as:
\begin{align}\label{FP}
	\underset{\boldsymbol\varphi}{\operatorname{max}}\  \frac{f(\boldsymbol\varphi )}{g(\boldsymbol\varphi )},\quad \text { s.t. }  \boldsymbol\varphi  \in \mathcal{S}, 
\end{align}
where $f(\boldsymbol\varphi )\geq 0$ is a nonnegative function, $g(\boldsymbol\varphi )>0$ is a positive function, and $\mathcal{S}$ is a nonempty constraint set.
The \ac{FP} problem~(\ref{FP}) can be transformed using Dinkelbach’s Transform~\cite{15} to a sequence of subproblems, where each subproblem is formulated as: 
\begin{align}\label{FP1}
	\underset{\boldsymbol\varphi}{\operatorname{max}}\   f(\boldsymbol\varphi )-\rho^{(t)} g(\boldsymbol\varphi),\quad
	\text { s.t. }  \boldsymbol\varphi  \in \mathcal{S},
\end{align}
where $t$ is the iteration index and $\rho$ is an introduced auxiliary variable, which is iteratively updated by $\rho^{(t)}=\frac{f(\boldsymbol\varphi ^{(t-1)})}{g(\boldsymbol\varphi ^{(t-1)})}$.
In~\cite{15a}, it is shown that for the case where the numerator and denominator of the fractional objective function are continuous and real-valued functions of $\boldsymbol\varphi  \in \mathcal{S}$, the globally optimal solution to~(\ref{FP}) can be obtained by iteratively solving the subproblems~(\ref{FP1}). In this work, we demonstrate that for problem~(\ref{eq:obj2}), where the numerator and denominator of the fractional objective function are real-valued but discrete functions, the globally optimal solution can also be found using the Dinkelbach’s Transform, followed by the \ac{SS} method for solving the subproblems.

Applying the Dinkelbach’s Transform to the reduced \ac{SINR} maximization problem~(\ref{eq:obj2}), the following subproblem needs to be solved in the $t$-th iteration: 
\begin{align}\label{OP1}
\underset{\varphi_1,...,\varphi_{N-1} }{\operatorname{max}}\   \mathbf{w}^{H}(\mathbf{B}(\theta)-\rho^{(t)}\mathbf{G})\mathbf{w}, \quad 
\text { s.t. } \varphi_n \in \mathcal{D}, \ \forall n,
\end{align}
where $\rho^{(t)}$ is given by: 
\begin{align}\label{updaterho}
\rho^{(t)}=\frac{(\mathbf{w}^{(t-1)})^{H}\mathbf{B}(\theta)\mathbf{w}^{(t-1)}}{(\mathbf{w}^{(t-1)})^{H}\mathbf{G}\mathbf{w}^{(t-1)}},
\end{align}
and $\mathbf{w}^{(t-1)}$ is the solution found in the $(t-1)$-th iteration. Denoting 
	$\mathbf{C}^{(t)}(\theta)=-(\mathbf{B}(\theta)-\rho^{(t)}\mathbf{G})\in \mathbb{C}^{N\times N}$,
the subproblem~(\ref{OP1}) can be rewritten as:
\begin{align}\label{SS1}
\underset{\varphi_1,...,\varphi_N }{\operatorname{min}}\   \mathbf{w}^{H}\mathbf{C}^{(t)}(\theta)\mathbf{w}, \ 
\text { s.t. } \varphi_n \in \mathcal{D}, \ \forall n.
\end{align}
\subsection{Discrete Codeword Search by the SS Method}\label{3B}

In this subsection, we present the \ac{SS} method, which adapts the idea of the \ac{CSD}~\cite{a10,aa10}, to solve the subproblem~(\ref{SS1}).
The studies in~\cite{14R,14RR} also employ \ac{SD} for discrete phase shift design. Whereas their approach utilizes the so-called \ac{RSD}~\cite{aa10}, which operates within a set of real numbers, we adopt the \ac{CSD} and perform the search directly within a set of complex numbers.

A prerequisite for applying the SS is that $\mathbf{C}^{(t)}(\theta)$ must be \ac{PD}, \tcc{as the Cholesky decomposition is performed on $\mathbf{C}^{(t)}(\theta)$. However, this condition is not necessarily guaranteed in our case.} To address this issue, we define $\gamma$ as the smallest eigenvalue of $\mathbf{C}^{(t)}(\theta)$ and define $\widehat{\mathbf{C}}^{(t)}(\theta)$ as: 
\begin{align}\label{PD}
	\widehat{\mathbf{C}}^{(t)}(\theta)=\mathbf{C}^{(t)}(\theta)+2|\gamma|\cdot\mathbf{I}_N.
\end{align}
This ensures that $\widehat{\mathbf{C}}^{(t)}(\theta)$ is always \ac{PD}.
Since $\mathbf{w}$ is \ac{CM}, we have: 
\begin{align}
\mathbf{w}^{H}\widehat{\mathbf{C}}^{(t)}(\theta)\mathbf{w}=\mathbf{w}^{H}\mathbf{C}^{(t)}(\theta)\mathbf{w}+2|\gamma|\cdot N.
\end{align}
Adding the constant term $2|\gamma|\cdot N$ to the objective function in~(\ref{SS1}) does not affect the optimization result. Therefore, the problem~(\ref{SS1}) is equivalent to: 
\begin{align}\label{SS}
	\vspace{-0.2mm}
\underset{\varphi_1,...,\varphi_N }{\operatorname{min}}\ \   \mathbf{w}^{H}\widehat{\mathbf{C}}^{(t)}(\theta)\mathbf{w}, \ 
\text { s.t. } \varphi_n \in \mathcal{D}, \ \forall n.
\vspace{-0.2mm}
\end{align}

\tcc{Note that the selection of $\gamma$ in Equation~\eqref{PD} has been discussed in~\cite{14RRR}, which shows that this parameter affects the algorithmic complexity, and that determining its optimal value in closed form is difficult. In our implementation, $\gamma$ is chosen as the smallest eigenvalue of $\mathbf{C}^{(t)}(\theta)$.}

Now the spherical search can be applied to solve the problem in~(\ref{SS}). First, we apply the Cholesky decomposition~\cite{14RRR,aa9} to $\widehat{\mathbf{C}}^{(t)}(\theta)$, obtaining an upper triangular matrix $\mathbf{U}^{(t)}(\theta)$ such that  $\widehat{\mathbf{C}}^{(t)}(\theta)=\mathbf{U}^{(t)}(\theta)^H\mathbf{U}^{(t)}(\theta)$. Next, we solve the following problem:
\begin{align} 
	\vspace{-0.5mm}
\underset{\varphi_1,...,\varphi_N }{\min}\    \left \| \mathbf{U}^{(t)}(\theta)\mathbf{w} \right \|^2_2,\  \text{s.t.}\   \varphi_n \in \mathcal{D},\  \forall n,
\vspace{-0.5mm}
\end{align}
which is essentially identical to~(\ref{SS}), and can be solved by adopting the idea of the \ac{CSD}, which was designed for MIMO detection applications for complex constellations. 
Based on the principle of the \ac{CSD}, the SS method searches only over the points that lie inside an $N$-dimensional hypersphere of radius $r$. This constraint can be expressed as: 
\begin{align} \label{eq:11}
	\vspace{-0.5mm}
\left \| \mathbf{U}^{(t)}(\theta)\mathbf{w} \right \|^2_2 \leq r^2.
\vspace{-0.5mm}
\end{align}
Let $\mathbf{p}=\mathbf{U}^{(t)}(\theta)\mathbf{w} \in \mathbb{C}^{N\times 1}$, and let $p_n$ denote the $n$-th entry of $\mathbf{p}$. Then, inequality~(\ref{eq:11}) can be written as: 
\begin{align}\label{23}
\sum_{n=1}^{N} |p_{n}|^2 \leq r^2.
\end{align}
Due to the upper triangular structure of $\mathbf{U}^{(t)}(\theta)$,
$p_n$ can be computed based on only the entries $w_j$ of $\mathbf{w}$, where $j = N, N-1, ..., n$, as follows:
\begin{align}\label{pn}
p_{n} =\mathrm{U}_{n,n}\left(w_{n}-\widehat{w}_{n}\right),
\end{align}
where $\widehat{w}_{n}$ is defined as:
\begin{align}\label{24}
	\widehat{w}_{n}=\left\{
	\begin{array}{lcl}
		0, & & {\text{for} \ n=N} 	\\
		-\sum_{j=n+1}^{N} \frac{\mathrm{U}_{n,j}}{\mathrm{U}_{n,n}}w_{j}, & &  {\text{for} \ n<N}	
	\end{array} \right.,
\end{align}
Here, $\mathrm{U}_{n,j}$ denotes the $(n,j)$-th entry of $\mathbf{U}^{(t)}(\theta)$, and $w_{j}$ the $j$-th entry of $\mathbf{w}$.
Since $|p_n|^2$ is positive, we have: 
\begin{align}\label{24a}
	\vspace{-0.5mm}
	\sum_{i=n}^{N} |p_{i}|^2 \leq r^2, \ \text{for} \  n=N,...,1.
	\vspace{-0.5mm}
\end{align} 
%\vspace{-1mm}
%In particular, for $n=N$, the inequality 
%$p_{N}^2\leq r^2$ must hold, which is equivalent to $\mathrm{U}_{N,N}^2w_N= \mathrm{U}_{N,N}^2\leq r^2$ and does not constitute a constraint for $w_N$ since we have fixed $\varphi_N=0$ (note that $w_{n}=e^{j \varphi_{n} }$).  
Based on~(\ref{pn}) and~(\ref{24a}), the search of $w_n$ can be conducted from $n=N-1$ down to $n=1$ as follows: First, initialize $n=N-1$. For $n=N-1$, inequality~(\ref{24a}) becomes:
\vspace{-1mm}
\begin{align}\label{27}
|p_{N-1}(w_{N-1})|^2 \leq r^2 -|p_{N}|^2,
\end{align}
which gives a condition for a feasible $w_{N-1}$ without violating~(\ref{eq:11}) or ~(\ref{23}) (note that $|p_N|^2 = \mathrm{U}_{N,N}^2$ since $w_N$ is fixed to be $1$). Later in this section, we will demonstrate that, based on the condition in~(\ref{27}), upper and lower bounds can be derived for feasible phases that can be assigned to $\varphi_{N-1}$ (note that $w_{n}=e^{j\varphi_n}$). For a chosen allowable value of $\varphi_{N-1}$, the inequality~(\ref{24a}), with $n=N-2$ substituted, becomes: 
\vspace{-1mm}
\begin{align}\label{28}
|p_{N-2}(w_{N-2})|^2 \leq r^2 -|p_{N}|^2-|p_{N-1}|^2,
\end{align}
which provides bounds on $\varphi_{N-2}$. Similarly, bounds on $\varphi_{n}$ for $n=N-3,...,1$, can be obtained, when the values for $\varphi_{n+1}, ..., \varphi_{N-1}$ have already been assigned.
\begin{algorithm}[t]
	%\vspace{5 mm}
	%\textsl{}\setstretch{1.8}
	\renewcommand{\algorithmicrequire}{\textbf{Input:}}
	\renewcommand{\algorithmicensure}{\textbf{Output:}}
	
	\caption{Discrete Codeword Search by SS}
	\label{alg1}
	
	\begin{algorithmic}[1]
		\REQUIRE{$\mathbf{U}^{(t)}(\theta)$, $N$, sphere radius $r^{(t)}$, step size $\Delta$}.
		\renewcommand{\algorithmicrequire}{\textbf{Initialize:}}
		%\State  \textbf{Initialization:} Set $w_N=1$, $r_N'=r$, $\mu_N=\mathrm{U}_{N,N}^2$, $\widehat{w}_N=0$, and $n=N-1$.
		\REQUIRE{$w_N=1$, $r=r^{(t)}$, $r_N'=r^{(t)}$, $p_N=\mathrm{U}_{N,N}$, $\widehat{w}_N=0$, and $n=N-1$.}
		\STATE (Bounds for $\varphi_n$) Compute $\widehat{w}_{n}$ and $r_n'$ based on~(\ref{24}) and ~(\ref{eq:12}), respectively. 
		Calculate $\widehat{\varphi}_{n}=arg(\widehat{w}_{n})$, and compute $\eta$ based on~(\ref{eq:16}).
		
		If $\eta>1,$ go to Step $3$. Else if $\eta<-1$, UB($n$) = $2\pi-\Delta$, $\varphi_{n}=-\Delta$, go to Step $2$. Else, UB($n$) = $\left\lfloor\left(\widehat{\varphi}_{n}+\cos ^{-1} \eta\right)\right\rfloor$, $\varphi_{n}=\left\lceil\left(\widehat{\varphi}_{n}-\cos ^{-1} \eta\right)\right\rceil-\Delta$, go to Step $2$.
		
		\STATE (Increase $\varphi_n$) $\varphi_n=\varphi_n+\Delta$. If $\varphi_n \leq$ UB($n$), then go to Step $4$. Else, go to Step $3$.
		\STATE (Increase $n$) $n=n+1$. If $n=N$, terminate. Else, go to Step $2$.
		\STATE (Decrease $n$) $p_n=\mathrm{U}_{n,n}(w_n-\widehat{w}_{n}).$
		If $n=1$,  then go to Step $5$. Else, $n=n-1$ and go to Step $1$.
		\STATE (Solution found) $r_{temp}=\sqrt{r_N'^2-r_1'^2+|p_1|^2}$. If $r_{temp}<r$, then let $r=r_{temp}$ and $\mathbf{w}_*=[e^{j \varphi_1 },...,e^{j \varphi_{N-1}},1 ] ^T$. Go to Step $2$.
		%\ENSURE  decomposed modes $ \left\{ {{s_k}\left( t \right)} \right\}$, $\left\{ {{\omega _k}\left( t \right)} \right\}$
		\ENSURE{$\mathbf{w}_*$}
	\end{algorithmic}  
\end{algorithm}
\addtolength{\topmargin}{0.05in}
\vspace{-0.5mm}
\begin{algorithm}[t]
	%\vspace{5 mm}
	%\textsl{}\setstretch{1.8}
	\renewcommand{\algorithmicrequire}{\textbf{Input:}}
	\renewcommand{\algorithmicensure}{\textbf{Output:}}
	\caption{Discrete Codeword Search by FP-SS}
	\label{alg2}
	
	\begin{algorithmic}[1]
		\REQUIRE{$\mathbf{w}^{(0)}$, $\mathbf{B}(\theta)$, $\mathbf{G}$, $N$, $\Delta$, $t_{\mathrm{max}}$}, $\varepsilon$.
		\renewcommand{\algorithmicrequire}{\textbf{Initialize:}}
		\REQUIRE{$t=0$, $\rho^{(0)}=0$.}
		%\State  Initialization: $t=0$, $\rho^{(0)}=\frac{(\mathbf{w}^{(0)})^{H}\mathbf{B}(\theta)\mathbf{w}^{(0)}}{(\mathbf{w}^{(0)})^{H}\mathbf{G}\mathbf{w}^{(0)}}$.
		\FOR{$t=1$ to $t_{\mathrm{max}}$}
		\STATE Calculate $\rho^{(t)}$ as in~(\ref{updaterho}). 
		If $|\rho^{(t)}-\rho^{(t-1)}|<\varepsilon$ : STOP. %$F(\rho^{(t)})<\varepsilon$
		\STATE Calculate $\mathbf{C}^{(t)}(\theta)=-(\mathbf{B}(\theta)-\rho^{(t)}\mathbf{G})$ and $\widehat{\mathbf{C}}^{(t)}(\theta)$ as in~(\ref{PD}). Obtain $\mathbf{U}^{(t)}(\theta)$ by the Cholesky decomposition of $\widehat{\mathbf{C}}^{(t)}(\theta)$, and the sphere radius $r^{(t)}=\sqrt{(\mathbf{w}^{(t-1)})^{H}\widehat{\mathbf{C}}^{(t)}(\theta)\mathbf{w}^{(t-1)}}$.
		
		\STATE $\mathbf{w}^{(t)}$ = SS($\mathbf{U}^{(t)}(\theta)$, $N$, $r^{(t)}$, $\Delta$).  \COMMENT{refer to Algorithm~\ref{alg1}}
		
		%\STATE $t\leftarrow t+1$.
		\ENDFOR
		\ENSURE{$\mathbf{w}^{(t)}$}.
		%\ENSURE  decomposed modes $ \left\{ {{s_k}\left( t \right)} \right\}$, $\left\{ {{\omega _k}\left( t \right)} \right\}$
	\end{algorithmic} 
	%\vspace{-0.5mm} 
\end{algorithm}
%\vspace{-0.1mm}

In the following, we describe how to obtain the upper and lower bounds for feasible phases that can be assigned to $\varphi_{N-1}$ based on condition~(\ref{27}), to $\varphi_{N-2}$ based on condition~(\ref{28}), and similarly for $n=N-3,...,1$. The set of codewords $e^{j\mathcal{D}}$ forms a constellation of $M$-PSK, where $M=2^b$ is the number of possible phases in the phase setting. Inspired by the \ac{CSD}~\cite{a10}, which gives boundaries for MIMO detection of $M$-PSK modulation, we can derive the lower and upper bounds for $\varphi_n$, i.e., the phase setting of antenna index $n$, in a similar way.
Define $r'_n$ as the new radius constraint for $\varphi_n$ due to the so far assigned phases for $\varphi_{n+1}, \varphi_{n+2}, ..., \varphi_{N}$, and it can be updated as:\vspace{-0.5mm}
\begin{align}
	\vspace{-0.3mm}
	%r_n'=r_{n+1}'^2-p_{n+1}^2,\label{eq:12}
	r_n'=\sqrt{r_{n+1}'^2-|p_{n+1}|^2},\label{eq:12}
	\vspace{-0.3mm}
\end{align}
where $r_N'^2=r^2$ is defined and $p_{n+1}$ is computed based on~(\ref{pn}). In the search conducted starting from antenna $N-1$ down to $1$, suppose $\varphi_{n+1}, ..., \varphi_{N-1}$ have already been assigned. \tcr{At antenna element $n$, the contribution to the sphere constraint must be within the remaining radius budget:}
\tcr{\begin{align}
		\mathrm{U}_{n,n}^2|w_n-\widehat{w}_{n}|^2 \leq r_n'^2,\label{eq:12A}
\end{align}}
\tcr{Based on Equation~(\ref{eq:12A}) and following the idea of the \ac{CSD}~\cite{a10}, a feasible $\varphi_n$ for antenna element $n$ should satisfy:}
\vspace{-1mm}
\begin{align}\label{eq:16}
	\cos \left(\varphi_n-\widehat{\varphi}_{n}\right) \geq \frac{1}{2  |\widehat{w}_{n}|}\left[1+|\widehat{w}_{n}|^{2}-\frac{{r^{\prime}}_{n}^{2}}{\mathrm{U}^{2}_{n, n}}\right] \overset{\text { def }}{=} \eta, 
	%\mathrm{U}_{n, n}{ }^{2}
\end{align}
where $\widehat{\varphi}_{n}= arg(\widehat{w}_{n})$, and $\widehat{w}_{n}$ is computed based on~(\ref{24}). 
%We define $r_N'^2=r^2$, $\widehat{w}_{N}=0$, $\mu_N=\mathrm{U}_{N,N}^2$, and the expressions~(\ref{eq:12})-(\ref{eq:14}) for $n=1,...,N-1$:
%\begin{align}
%r_n'^2=r_{n+1}'^2-\mu_{n+1},\label{eq:12}\\
%\widehat{w}_{n}=-\sum_{j=n+1}^{N} %\frac{\mathrm{U}_{n,j}}{\mathrm{U}_{n,n}}w_{j},\label{eq:13}\\
%\mu_n=\mathrm{U}_{n,n}^2|w_n-\widehat{w}_{n}|^2.\label{eq:14}
%\end{align}

\begin{figure}[t]
	\centering
	\includegraphics[trim={0cm 0cm 0cm 0cm},clip,width=0.45\textwidth]{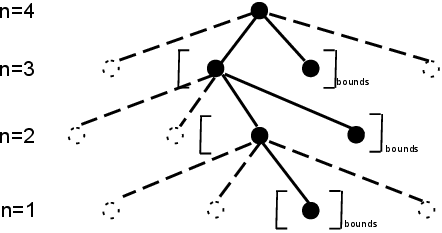}
	\caption{Tree search example with $2$-bit phase quantization and $4$ RX antennas.}
	\label{fig2}
	\vspace{-2mm}
\end{figure}
If $\eta >1$, then at this level no codeword within the set $e^{j\mathcal{D}}$ is feasible. If $\eta <-1$, the entire set $e^{j\mathcal{D}}$ is feasible. For $-1\leq \eta \leq 1$, the allowable phase range for $\varphi_n$ is given by:
%\vspace{-1mm}
\begin{align}\label{eq:17}
\left\lceil\left(\widehat{\varphi}_{n}-\cos ^{-1} \eta\right)\right\rceil \leq \varphi_n \leq\left\lfloor\left(\widehat{\varphi}_{n}+\cos ^{-1} \eta\right)\right\rfloor,
\end{align}
where the functions $\left\lceil\cdot \right\rceil$ and $\left\lfloor\cdot \right\rfloor$ refer to the next greater/smaller phase $\varphi$ in $\mathcal{D}$.

The full SS procedure can be illustrated as a search tree from $n=N$ down to $n=1$, where each path through the tree corresponds to a possible codeword, as shown in Fig.~\ref{fig2}. The search with the boundary control can be viewed as a pruning algorithm on this tree where at some depth the subtrees can be rejected based on violation of the constraint given by~(\ref{eq:16}). SS traces down branches of the search tree until no more allowable points exist in the current level. At this point, it backtracks and proceeds down a different branch. 

Algorithm~\ref{alg1} describes the details of the proposed discrete codeword search using the SS method to solve the subproblem~(\ref{OP1}). The SS method is further integrated into the FP iterations, named as FP-SS, which is shown in Algorithm~\ref{alg2}. 
Note that in Algorithm~\ref{alg2}, the initial codeword could be set to a random discrete phase vector or by a hard-quantized version of a continuous phase vector found by a low-complexity continuous optimization method, such as the MVDR method~\cite{10}.
%\vspace{-2mm}
\subsection{Optimality and Convergence}\label{optimality}
Let $f(\mathbf{w})=\mathbf{w}^{H}\mathbf{B}(\theta)\mathbf{w}$, $g(\mathbf{w})=\mathbf{w}^{H}\mathbf{G}\mathbf{w}$, the feasible region $\mathcal{S}$: $\{\mathbf{w} \mid \mathbf{w}=[e^{j \varphi_1 },...,e^{j \varphi_{N-1}},1 ] ^T, \text{with}\  \varphi_1,...,\varphi_{N-1} \in \mathcal{D}\}$, and $F(\rho)=\underset{\mathbf{w}}{\max}  \{f(\mathbf{w})-\rho g(\mathbf{w})\mid\mathbf{w}\in \mathcal{S}\}$.

\textbf{Proposition $1$.} (Optimality) $F(\rho^*)=\underset{\mathbf{w}}{\max}\   \{f(\mathbf{w})-\rho^*g(\mathbf{w})\mid\mathbf{w}\in \mathcal{S}\}=f(\mathbf{w}^*)-\rho^*g(\mathbf{w}^*)=0$  if and only if $\rho^*=\frac{f(\mathbf{w}^*)}{g(\mathbf{w}^*)}=\underset{\mathbf{w}}{\max}  \{\frac{f(\mathbf{w})}{g(\mathbf{w})}\mid\mathbf{w}\in \mathcal{S}\}.$

\textit{Proof}: See Appendix~\ref{appendix1}.

\textbf{Proposition $2$.} (Convergence) Dinkelbach’s algorithm converges superlinearly for the problem in~(\ref{eq:obj2}). 

\textit{Proof}: See Appendix~\ref{appendix1}.

As shown in Subsection~\ref{3B}, the subproblems $\underset{\mathbf{w}}{\max}\   \{f(\mathbf{w})-\rho g(\mathbf{w})\mid\mathbf{w}\in \mathcal{S}\}$ can be solved by the \ac{SS} method, meaning that the optimal solution of the subproblems can be found using the \ac{SS} method. Based on Proposition $1$ and Proposition $2$, we can solve the IQFP problem in~(\ref{eq:obj2}) by solving a sequence of subproblems~(\ref{OP1}) using the SS method, and eventually obtain the globally optimal solution to the IQFP problem in~(\ref{eq:obj2}).
\section{Discrete TX Codebook Design with a Fixed RX Codeword}\label{TX}
In this section, we assume that the RX codeword $\mathbf{w}$ is fixed as $\mathbf{w}=\mathbf{w}_{0}$, which is properly selected, for instance, as the quantized version of the steering vector pointing to the sensing direction. The task is then to maximize the receiver SINR \ac{w.r.t.} $\mathbf{v}$. Analogous to Section~\ref{RX}, we show that this simplified problem can again be formulated as an \ac{IQFP} problem, and then transformed into a sequence of subproblems using Dinkelbach's algorithm. However, unlike in Section~\ref{RX}, the subproblems now have an inequality constraint and cannot be solved using the previously proposed \ac{SS} method. To address this, we further generalize the \ac{SS} method and propose the \ac{CSS} method, which considers the extra inequality constraint.
Using the FP-CSS method, which combines the \ac{FP} technique and the \ac{CSS} method, we can find the globally optimal solution to the simplified TX codebook design problem. 

Denote $\tilde{\mathbf{b}}^{H}(\theta)=\mathbf{w}_{0}^{H}\mathbf{a}_r(\theta)\mathbf{a}_t^{H}(\theta)\in \mathbb{C}^{1\times M}$ and $\tilde{\mathbf{g}}^{H}_{\mathrm{SI}}=\mathbf{w}_{0}^{H}\mathbf{H}_{\mathrm{SI}}\in \mathbb{C}^{1\times M}$. 
%\sout{Note that $\mathbf{b}(\theta)$ is $\mathbf{a}_{r}(\theta)$ scaled by a complex number.}
The objective function in~(\ref{op}), $q_{\theta}(\mathbf{w}=\mathbf{w}_0,\mathbf{v})$, can be reduced to: 
\begin{align}\tilde{q}_{\theta}(\mathbf{v})
	&=\frac{\left | \mathbf{v}^{H}\tilde{\mathbf{b}}(\theta) \right |^{2}}{\left | \mathbf{v}^{H} \tilde{\mathbf{g}}_{\mathrm{SI}} \right |^{2}+ \frac{N}{P_{t}}\sigma_z^{2}}\\
	&=\frac{\left | \mathbf{v}^{H} \tilde{\mathbf{b}}(\theta) \right |^{2}}{\mathbf{v}^{H}(\widetilde{\mathbf{G}}_{\mathrm{SI}}+\frac{N\sigma_z^{2}}{MP_{t}}\mathbf{I}_M)\mathbf{v}},
\end{align} 
where $\widetilde{\mathbf{G}}_{\mathrm{SI}}=\tilde{\mathbf{g}}_{\mathrm{SI}}\tilde{\mathbf{g}}_{\mathrm{SI}}^{H}\in \mathbb{C}^{M\times M}$. We further define $\widetilde{\mathbf{G}}=\widetilde{\mathbf{G}}_{\mathrm{SI}}+\frac{N\sigma_z^{2}}{MP_{t}}\mathbf{I}_M\in \mathbb{C}^{M\times M}$ and $\widetilde{\mathbf{B}}(\theta)= \tilde{\mathbf{b}}(\theta)\tilde{\mathbf{b}}(\theta)^{H}\in \mathbb{C}^{M\times M}$. Then, the reduced \ac{SINR} maximization problem at $\theta$ \ac{w.r.t.} the TX codeword $\mathbf{v}$ is:
%\vspace{-0.5mm}
 \begin{equation}
	\begin{split}\label{eq:obj3}
	\underset{\psi_1,...,\psi_{M-1} }{\max} \frac{\mathbf{v}^{H}\widetilde{\mathbf{B}}(\theta)\mathbf{v}}{\mathbf{v}^{H}\widetilde{\mathbf{G}}\mathbf{v}}, \  \text { s.t. } \psi_1,...,\psi_{M-1} \in \mathcal{D},\\
	\text{s.t.}\ \  |\mathbf{v}^{H}\mathbf{a}_{t}(\theta_{c})|^2 \geq c^2.
 	\end{split}
\end{equation}
Compared to Section~\ref{RX}, one extra inequality constraint needs to be considered. In the following we will show that this constraint can provide another feasible phase range for each antenna element, in addition to the existing feasible range identified in the \ac{SS} method. We denote the feasible region of $\mathbf{v}$ as $\widetilde{S}$: $\widetilde{S}=\{\mathbf{v} \mid \mathbf{v}=[e^{j \psi_1 },...,e^{j \psi_{M-1}},1 ] ^T, \text{with}\  \psi_1,...,\psi_{M-1} \in \mathcal{D}, |\mathbf{v}^{H}\mathbf{a}_{t}(\theta_{c})|^2 \geq c^2\}$.
\subsection{Problem Transformation by the FP method}\label{range1}
Analogous to Section~\ref{RX}, Dinkelbach's algorithm can be applied to the objective function in~(\ref{eq:obj3}), and the problem~(\ref{eq:obj3}) can be solved iteratively. In the $t$-th iteration, the following subproblem needs to be solved:
 \begin{equation}
	\begin{split}\label{eq:obj4}
		\underset{\psi_1,...,\psi_{M-1} }{\max} \mathbf{v}^{H}(\widetilde{\mathbf{B}}(\theta)-\tilde{\rho}^{(t)}\widetilde{\mathbf{G}})\mathbf{v}, \  \text { s.t. } \mathbf{v} \in \widetilde{S},
	\end{split}
\end{equation}
where $\tilde{\rho}^{(t)}$ is given by: 
\begin{align}\label{updaterho2}
	\tilde{\rho}^{(t)}=\frac{(\mathbf{v}^{(t-1)})^{H}\widetilde{\mathbf{B}}(\theta)\mathbf{v}^{(t-1)}}{(\mathbf{v}^{(t-1)})^{H}\widetilde{\mathbf{G}}\mathbf{v}^{(t-1)}}.
\end{align}
Here, $\mathbf{v}^{(t-1)}$ is the solution found in the $(t-1)$-th iteration. Denote 
$\mathbf{E}^{(t)}(\theta)=-(\widetilde{\mathbf{B}}(\theta)-\tilde{\rho}^{(t)}\widetilde{\mathbf{G}})\in \mathbb{C}^{M\times M}$, and define 
\begin{align}\label{PD2}
	\widehat{\mathbf{E}}^{(t)}(\theta)=\mathbf{E}^{(t)}(\theta)+2|\tilde{\gamma}|\cdot\mathbf{I}_M,
\end{align}
where $\tilde{\gamma}$ is the smallest eigenvalue of $\mathbf{E}^{(t)}(\theta)$. This ensures that $\widehat{\mathbf{E}}^{(t)}(\theta)$ is \ac{PD}.
Further, apply the Cholesky decomposition to $\widehat{\mathbf{E}}^{(t)}(\theta)$ and obtain the upper triangular matrix $\widetilde{\mathbf{U}}^{(t)}(\theta)$ from  $\widehat{\mathbf{E}}^{(t)}(\theta)=\widetilde{\mathbf{U}}^{(t)}(\theta)^H\widetilde{\mathbf{U}}^{(t)}(\theta)$. The subproblem~(\ref{eq:obj4}) is then equivalent to:
\begin{equation}\label{eq:obj5}
	\begin{split}
	\underset{\psi_1,...,\psi_{M-1} }{\min}\    \left \| \widetilde{\mathbf{U}}^{(t)}(\theta)\mathbf{v} \right \|^2_2, \  \text { s.t. } \mathbf{v} \in \widetilde{S}.
	\end{split}
\end{equation}
\subsection{Discrete Codeword Search by the \ac{CSS} Method}\label{range2}
Denote $\mathbf{A}_c=\mathbf{a}(\theta_{c})\mathbf{a}^{H}(\theta_{c})\in \mathbb{C}^{M\times M}$. Then, the inequality of the communication constraint can be reformulated as: 
\begin{gather}
	\mathbf{v}^{H}\mathbf{A}_c\mathbf{v}\geq c^2\\
\Leftrightarrow\  -\mathbf{v}^{H}\mathbf{A}_c\mathbf{v}+\alpha\cdot\mathbf{v}^{H}\mathbf{v}\leq -c^2+\alpha\cdot\mathbf{v}^{H}\mathbf{v}\label{a1}\\
\Leftrightarrow\  \mathbf{v}^{H}(-\mathbf{A}_c+\alpha\cdot\mathbf{I}_M)\mathbf{v}\leq -c^2+\alpha\cdot M\label{a2}\\
\Leftrightarrow\  \mathbf{v}^{H}\widehat{\mathbf{A}}_c\mathbf{v}\leq \widehat{c}^2,
\end{gather}
where $\widehat{\mathbf{A}}_c$ and $\widehat{c}$ are defined as $\widehat{\mathbf{A}}_c=-\mathbf{A}_c+\alpha\cdot\mathbf{I}_M$ and $\widehat{c}^2=-c^2+\alpha\cdot M$, respectively. From~(\ref{a1}) to~(\ref{a2}), the CM property of $\mathbf{v}$ is used. $\widehat{\mathbf{A}}_c$ can be guaranteed to be a PD matrix through the selection of $\alpha$.

To 
$\widehat{\mathbf{A}}_c$ we again apply the Cholesky decomposition, and obtain another upper triangular matrix $\mathbf{U}_c$ from $\widehat{\mathbf{A}}_c=\mathbf{U}_c^H\mathbf{U}_c$. The communication constraint inequality in~(\ref{eq:obj3}) is then equivalent to:
\begin{align} \label{constraint}
\left \| \mathbf{U}_c\mathbf{v} \right \|^2_2\leq \widehat{c}^2.
\end{align}
The problem~(\ref{eq:obj5}) can now be reformulated as:
\begin{equation}\label{eq:obj6}
	\begin{split}
		\underset{\psi_1,...,\psi_{M-1} }{\min}\    \left \| \widetilde{\mathbf{U}}^{(t)}(\theta)\mathbf{v} \right \|^2_2, \  \text { s.t. } \psi_1,...,\psi_{M-1} \in \mathcal{D},\\
		\text{s.t.}\ \  \left \| \mathbf{U}_c\mathbf{v} \right \|^2_2\leq \widehat{c}^2.
	\end{split}
\end{equation}
To solve the problem~(\ref{eq:obj6}), we can combine the constraint~(\ref{constraint}) with the \ac{SS} procedure described in Subsection~\ref{3B}. We set a proper initial radius $\tilde{r}$ for $\left \| \widetilde{\mathbf{U}}^{(t)}(\theta)\mathbf{v} \right \|$, and search only over those points $\mathbf{v}$ that lie within the $M$-dimensional hypersphere of radius $\tilde{r}$, as defined by the following constraint for the objective function in~(\ref{eq:obj6}):
%the following constraint for the objective function in~(\ref{eq:obj6}): 
\begin{align} \label{eq:13}
	\left \| \widetilde{\mathbf{U}}^{(t)}(\theta)\mathbf{v} \right \|^2_2 \leq \tilde{r}^2,
\end{align}
while simultaneously satisfying the communication constraint~(\ref{constraint}).

For antenna element $m$, supposing that $\psi_{m+1},...,\psi_{M}$ are already assigned, we can obtain an allowable range for $\psi_{m}$ based on~(\ref{eq:13}), and another allowable range for $\psi_{m}$ based on constraint~(\ref{constraint}). Taking the intersection of the two ranges, we can exclude the choices for $\psi_{m}$ that violate the constraint~(\ref{constraint}) or~(\ref{eq:13}), and get the final allowable range for $\psi_{m}$. 
\begin{algorithm}[t]
	%\vspace{5 mm}
	%\textsl{}\setstretch{1.8}
	\renewcommand{\algorithmicrequire}{\textbf{Input:}}
	\renewcommand{\algorithmicensure}{\textbf{Output:}}
	
	\caption{Discrete Codeword Search by CSS}
	\label{alg3}
	
	\begin{algorithmic}[1]
		\REQUIRE{$\widetilde{\mathbf{U}}^{(t)}(\theta)$, $\mathbf{U}_c$, $M$, sphere radius $\tilde{r}^{(t)}$, $\widehat{c}$, step size $\Delta$}.
		\renewcommand{\algorithmicrequire}{\textbf{Initialize:}}
		%\State  \textbf{Initialization:} Set $w_N=1$, $r_N'=r$, $\mu_N=\mathrm{U}_{N,N}^2$, $\widehat{w}_N=0$, and $n=N-1$.
		\REQUIRE{$v_M=1$, $\tilde{r}=\tilde{r}^{(t)}$, $\tilde{r}_M'=\tilde{r}^{(t)}$, $c_M'=\widehat{c}$, $\tilde{p}_M=\widetilde{\mathrm{U}}_{M,M}$,  $\tilde{p}_M'=\mathbf{U}^c_{M,M}$, $\widehat{v}_M=0$, and $m=M-1$.}
		\STATE (Bounds for $\psi_m$) Compute $\widehat{v}_m$ and $\tilde{r}_m'$ based on~(\ref{46}) and (\ref{48}), respectively,
		calculate $\widehat{\psi}_{m}=arg(\widehat{v}_{m})$. Then compute $\tilde{\eta}_1$ based on~(\ref{49}).
		
		Compute $\widehat{v}_{m}'$ and $c_m'$ based on~(\ref{50}) and (\ref{52}), respectively,
		calculate $\widehat{\psi}_{m}'=arg(\widehat{v}_{m}')$, compute $\tilde{\eta}_2$ based on~(\ref{53}).
		
		If $\tilde{\eta}_1>1$ or $\tilde{\eta}_2>1,$ go to Step $3$. Else, compute $InterSet(m,:)$, the intersection of the two preliminary allowable ranges, based on Appendix~\ref{appendix2}, set $i=0$, and $NoE(m)$ as the number of elements in $InterSet(m,:)$, go to Step~$2$. 
		
		\STATE (Increase index $i$) $i=i+1$. If $i \leq NoE(m)$, then set $\psi_m=InterSet(m,i)$ and go to Step $4$. Else, go to Step~$3$. 
		\STATE (Increase $m$) $m=m+1$. If $m=M$, terminate. Else, go to Step $2$.
		\STATE (Decrease $m$) $\tilde{p}_m=\widetilde{\mathrm{U}}_{m,m}(v_m-\widehat{v}_{m}).$
		If $m=1$,  then go to Step $5$. Else, $m=m-1$ and go to Step $1$.
		\STATE (Solution found) $\tilde{r}_{temp}=\sqrt{\tilde{r}_M'^2-\tilde{r}_1'^2+|\tilde{p}_1|^2}$. If $\tilde{r}_{temp}<\tilde{r}$, then let $\tilde{r}=\tilde{r}_{temp}$ and $\mathbf{v}_*=[e^{j \psi_1 },...,e^{j \psi_{M-1}},1 ] ^T$. Go to Step $2$.
		%\ENSURE  decomposed modes $ \left\{ {{s_k}\left( t \right)} \right\}$, $\left\{ {{\omega _k}\left( t \right)} \right\}$
		\ENSURE{$\mathbf{v}_*$}
	\end{algorithmic}  
	
\end{algorithm}

Define  $\tilde{r}_M'^2= \tilde{r}^2$, $\widehat{v}_M = 0$, $\tilde{p}_{M} = \widetilde{\mathrm{U}}_{M,M}$, where $\widetilde{\mathrm{U}}_{m,j}$ is the $m,j$-th element of matrix $\widetilde{\mathrm{U}}$. Further define the following equations for $m = 1, ..., M - 1$:
\begin{gather} 
		\widehat{v}_{m}=-\sum_{j=m+1}^{M}\label{46} \frac{\widetilde{\mathrm{U}}_{m,j}}{\widetilde{\mathrm{U}}_{m,m}}v_{j},\\
	\tilde{p}_{m} =\widetilde{\mathrm{U}}_{m,m}\left(v_{m}-\widehat{v}_{m}\right),\\
	\tilde{r}_m'= \sqrt{\tilde{r}_{m+1}'^2-|\tilde{p}_{m+1}|^2}.\label{48}
\end{gather}
Based on these equations, we can compute $\tilde{\eta}_1$:
\begin{gather}\label{49}
	\tilde{\eta}_1=\frac{1}{2  |\widehat{v}_{m}|}\left[1+|\widehat{v}_{m}|^{2}-\frac{\tilde{r}_{m}'^{2}}{\widetilde{\mathrm{U}}^{2}_{m, m}}\right]. 
\end{gather}
Similarly, we define $c_M'^2= \widehat{c}^2$, $\widehat{v}_M' = 0$, $\tilde{p}_{M}' = \mathrm{U}_{M,M}^c$, where $\mathrm{U}_{m,j}^c$ is the $m,j$-th element of matrix $\mathrm{U}_c$. Further define the following equations for $m = 1, ..., M - 1$:
\begin{gather} 
	\widehat{v}_{m}'=-\sum_{j=m+1}^{M} \frac{\mathrm{U}_{m,j}^c}{\mathrm{U}_{m,m}^c}v_{j},\label{50}\\
	\tilde{p}_{m}' =\mathrm{U}_{m,m}^c\left(v_{m}-\widehat{v}_{m}'\right),\\
	c_m'= \sqrt{c_{m+1}'^2-|\tilde{p}_{m+1}'|^2}.\label{52}
\end{gather}
Based on Equations~(\ref{50})--(\ref{52}), we can again compute $\tilde{\eta}_2$:
\begin{gather}\label{53}
	\tilde{\eta}_2=\frac{1}{2  |\widehat{v}_{m}'|}\left[1+|\widehat{v}_{m}'|^{2}-\frac{c_{m}'^{2}}{(\mathrm{U}_{m,m}^c)^2}\right]. 
\end{gather}
\begin{figure}[t]
	\centering
	\includegraphics[trim={0cm 0cm 0cm 0cm},clip,width=0.4\textwidth]{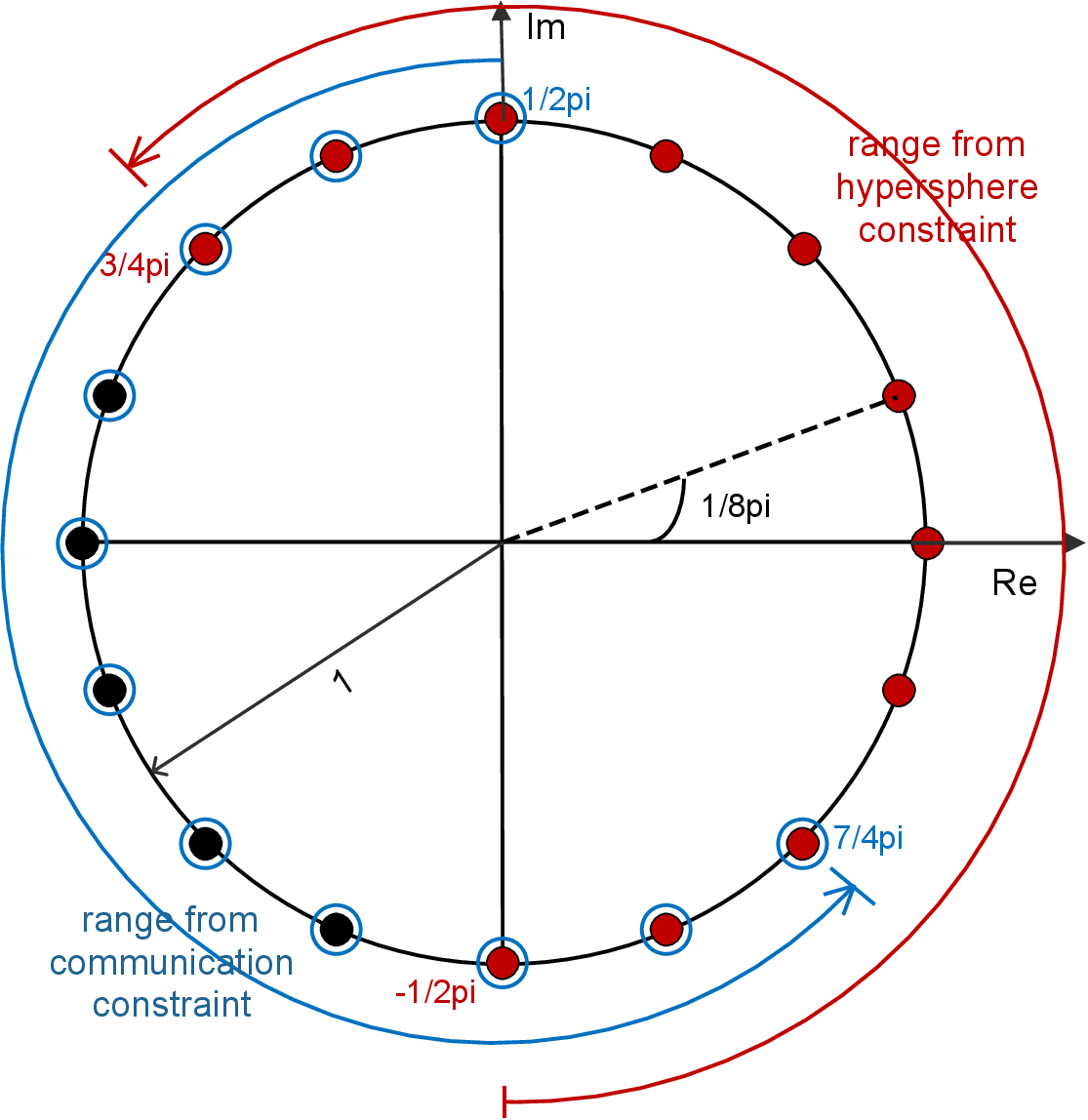}
	\caption{An example illustrating the intersection of two allowable ranges before normalization is presented. The intersection can result in either a single interval or two separate intervals. In this case, the intersection forms two separate intervals. Specifically, the intersection of the range $[-\frac{1}{2}\pi, \frac{3}{4}\pi]$ from the hypersphere constraint as in~(\ref{eq:13}) (denoted by red points), and the range $[\frac{1}{2}\pi, \frac{7}{4}\pi]$ from the communication constraint as in~(\ref{constraint}) (denoted by points surrounded by a blue circle), is shown. The final intersection consists of two separate intervals, with the points marked in red and encircled. For this example, a $4$-bit phase shift resolution is assumed.}
	\label{fig3}
\end{figure}
\begin{algorithm}[t]
	%\vspace{5 mm}
	%\textsl{}\setstretch{1.8}
	\renewcommand{\algorithmicrequire}{\textbf{Input:}}
	\renewcommand{\algorithmicensure}{\textbf{Output:}}
	\caption{Discrete Codeword Search by FP-CSS}
	\label{alg4}
	
	\begin{algorithmic}[1]
		\REQUIRE{$\mathbf{v}^{(0)}$, $\widetilde{\mathbf{B}}(\theta)$, $\widetilde{\mathbf{G}}$, $\mathbf{A}_c$, $c$, $M$, $\Delta$, $\tilde{t}_{\mathrm{max}}$}, $\tilde{\varepsilon}$.
		\renewcommand{\algorithmicrequire}{\textbf{Initialize:}}
		\REQUIRE{$t=0$, $\tilde{\rho}^{(0)}=0$.}
		%\State  Initialization: $t=0$, $\rho^{(0)}=\frac{(\mathbf{w}^{(0)})^{H}\mathbf{B}(\theta)\mathbf{w}^{(0)}}{(\mathbf{w}^{(0)})^{H}\mathbf{G}\mathbf{w}^{(0)}}$.
		\STATE Obtain $\widehat{\mathbf{A}}_c$ and $\widehat{c}$. Obtain $\mathbf{U}_c$ by the Cholesky decomposition of $\widehat{\mathbf{A}}_c$. 
		\FOR{$t=1$ to $\tilde{t}_{\mathrm{max}}$}
		\STATE Calculate $\tilde{\rho}^{(t)}$ as in~(\ref{updaterho2}). 
		%If $\tilde{F}(\tilde{\rho}^{(t)})<\tilde{\varepsilon}$: STOP.
		If  $|\tilde{\rho}^{(t)}-\tilde{\rho}^{(t-1)}|<\tilde{\varepsilon}$: STOP.
		\STATE Calculate $\mathbf{E}^{(t)}(\theta)=-(\widetilde{\mathbf{B}}(\theta)-\tilde{\rho}^{(t)}\widetilde{\mathbf{G}})$ and $\widehat{\mathbf{E}}^{(t)}(\theta)$ as in~(\ref{PD2}). Obtain $\widetilde{\mathbf{U}}^{(t)}(\theta)$ by the Cholesky decomposition of $\widehat{\mathbf{E}}^{(t)}(\theta)$, and the sphere radius $\tilde{r}^{(t)}=\sqrt{(\mathbf{v}^{(t-1)})^{H}\widehat{\mathbf{E}}^{(t)}(\theta)\mathbf{v}^{(t-1)}}$.
		
		\STATE $\mathbf{v}^{(t)} =$ CSS($\widetilde{\mathbf{U}}^{(t)}(\theta)$, $\mathbf{U}_c$, $M$, $\tilde{r}^{(t)}$, $\widehat{c}$, $\Delta$).  \COMMENT{refer to Algorithm~\ref{alg3}}
		
		%\STATE $t\leftarrow t+1$.
		\ENDFOR
		\ENSURE{$\mathbf{v}^{(t)}$}.
		%\ENSURE  decomposed modes $ \left\{ {{s_k}\left( t \right)} \right\}$, $\left\{ {{\omega _k}\left( t \right)} \right\}$
	\end{algorithmic}  
\end{algorithm}
Based on the calculated values of $\tilde{\eta}_1$ and $\tilde{\eta}_2$, we can determine two corresponding allowable ranges for $\psi_m$, and ultimately obtain the intersection of these two ranges. An example of the intersection of two allowable ranges is shown in Fig.~\ref{fig3}. 
The process of obtaining the intersection of the ranges based on $\tilde{\eta}_1$ and $\tilde{\eta}_2$ is presented in Appendix~\ref{appendix2}.
%In Appendix~\ref{appendix2} we present how to get the intersection of the ranges based on $\tilde{\eta}_1$ and $\tilde{\eta}_2$. 
With the boundary control, the CSS procedure can be performed in the same way as the SS procedure, namely can be illustrated as a search tree from $m=M$ down to $m=1$. Finally, the CSS method, as described in Algorithm~\ref{alg3}, is integrated into the FP iterations, named as FP-CSS, which is described in Algorithm~\ref{alg4}. It is worth mentioning that in Algorithm~\ref{alg4}, the initial codeword should be set to a discrete phase vector that satisfies the communication constraint, ensuring that the algorithm produces a valid solution. A possible choice for this initial codeword could be the hard-quantized version of the steering vector corresponding to the communication direction.
%In Algorithm~\ref{alg3}, $\tilde{F}(\tilde{\rho})$ is defined as $\tilde{F}(\tilde{\rho})= \widetilde{\mathbf{B}}(\theta)-\tilde{\rho}^{(t)}\widetilde{\mathbf{G}}$. 
Similar to the FP-SS method, the propositions for optimality and convergence also apply to the FP-CSS method, and the proof follows the same steps.
%\begin{algorithm}[t]
	%\vspace{5 mm}
	%\textsl{}\setstretch{1.8}
%	\renewcommand{\algorithmicrequire}{\textbf{Input:}}
%	\renewcommand{\algorithmicensure}{\textbf{Output:}}
%	\caption{Intersection of two allowble ranges}
%	\label{alg3}
	
%	\begin{algorithmic}[1]
%		\REQUIRE{$\tilde{\eta}_1$, $\tilde{\eta}_2$.}
%		\renewcommand{\algorithmicrequire}{\textbf{Initialize:}}
%		\REQUIRE{$t=0$, $\rho^{(0)}=\frac{(\mathbf{w}^{(0)})^{H}\mathbf{B}(\theta)\mathbf{w}^{(0)}}{(\mathbf{w}^{(0)})^{H}\mathbf{G}\mathbf{w}^{(0)}}$.}
%		%\State  Initialization: $t=0$, $\rho^{(0)}=\frac{(\mathbf{w}^{(0)})^{H}\mathbf{B}(\theta)\mathbf{w}^{(0)}}{(\mathbf{w}^{(0)})^{H}\mathbf{G}\mathbf{w}^{(0)}}$.
%		\STATE If $\tilde{\eta}_1\leq -1$ and $\tilde{\eta}_2\leq -1$: LB1 = $0$, UB1 = $2\pi-\Delta$.
%		\STATE Calculate $\mathbf{C}^{(t)}(\theta)=-(\mathbf{B}(\theta)-\rho^{(t)}\mathbf{G})$ and $\widehat{\mathbf{C}}^{(t)}(\theta)$ as in~(\ref{PD}). Obtain $\mathbf{U}^{(t)}(\theta)$ by the Cholesky decomposition of $\widehat{\mathbf{C}}^{(t)}(\theta)$, and the sphere radius $r^{(t)}=\sqrt{(\mathbf{w}^{(t-1)})^{H}\widehat{\mathbf{C}}^{(t)}(\theta)\mathbf{w}^{(t-1)}}$.
		
%		\STATE $\mathbf{w}^{(t)}$ = SS($\mathbf{U}^{(t)}(\theta)$, $N$, $r^{(t)}$, $\Delta$).  \COMMENT{refer to Algorithm~\ref{alg1}}
		
		%\STATE $t\leftarrow t+1$.
%		\ENSURE{LB1, UB1, LB2, UB2}.
		%\ENSURE  decomposed modes $ \left\{ {{s_k}\left( t \right)} \right\}$, $\left\{ {{\omega _k}\left( t \right)} \right\}$
%	\end{algorithmic}  
%\end{algorithm}

\section{Joint Discrete Codebook Design by Alternating Optimization}\label{joint}
\subsection{Joint Codebook Design via Alternating Optimization}
Combining the methods from Section~\ref{RX} and~\ref{TX}, we can jointly optimize the RX and TX codebooks by alternately updating them. As described in Algorithm~\ref{alg6}, the process begins by selecting an appropriate initialization for TX codeword $\mathbf{v}$, such as the quantized DFT beamforming vector toward the given communication direction $\theta_c$. With the initialized $\mathbf{v}$ fixed, update the RX codeword $\mathbf{w}$ by solving the \ac{OP} in~(\ref{eq:obj2}) using the FP-SS method, as described in Section~\ref{RX}. Next, fix the obtained $\mathbf{w}$ and update $\mathbf{v}$ by solving the \ac{OP} in~(\ref{eq:obj3}) based on the FP-CSS method, as explained in Section~\ref{TX}. Subsequently, fix the updated $\mathbf{v}$ and re-optimize $\mathbf{w}$. Repeat this alternating process until the stopping criterion is met. Note that the alternating optimization can also be initiated by starting with a properly initialized $\mathbf{w}$ and updating $\mathbf{v}$ first.

\subsection{Initialization Strategy}

\tcc{In the proposed alternating optimization based joint codebook design algorithm, the initial TX codeword $\mathbf{v}^{(0)}$ is chosen as the quantized DFT beamforming vector pointing toward the communication direction $\theta_c$. The initial RX codeword $\mathbf{w}^{(0)}$ is obtained using the MVDR-CM-HQ method, i.e., the hard-quantized and CM version of the classical MVDR beamformer.
It is worth noting that the initialization of $\mathbf{w}^{(0)}$ does not affect the final performance of the proposed algorithm. In the FP-SS procedure, $\mathbf{w}^{(0)}$ only determines the initial search radius in the first SS step; after that, FP-SS computes the optimal discrete update independently of the initial choice. Therefore, replacing $\mathbf{w}^{(0)}$ with other feasible  quantized vectors does not change the final converged solution.}
\begin{algorithm}%[t]
	%\scriptsize
	%\textsl{}\setstretch{1.8}
	\renewcommand{\algorithmicrequire}{\textbf{Input:}}
	\renewcommand{\algorithmicensure}{\textbf{Output:}}
	\caption{Joint Codebook Design by Alternating Optimization}
	\label{alg6}
	\begin{algorithmic}[1]
		\STATE Initialize $\mathbf{v}^{(0)}$, $\mathbf{w}^{(0)}$, set $t=1$.
		\WHILE{Stopping criterion not met}
		
		\STATE Fix $\mathbf{v}^{(t-1)}$ and obtain $\mathbf{w}^{(t)}$ based on the FP-SS method.	
		\STATE  Fix $\mathbf{w}^{(t)}$ and obtain $\mathbf{v}^{(t)}$ based on the FP-CSS method.	 
		\STATE $t\leftarrow t+1$.
		\ENDWHILE
		\ENSURE{$\mathbf{w}^{(t)}$}, {$\mathbf{v}^{(t)}$}.
	\end{algorithmic}  
\end{algorithm}  
\subsection{Convergence Analysis}
\tcc{Each FP-SS / FP-CSS update monotonically increases (or maintains) the objective value, based on the demonstrated optimality of FP-SS / FP-CSS.
At the same time, the overall objective is upper-bounded due to the discrete CM codebook constraint, which restricts all beamforming vectors to a finite and compact feasible set.
As a result, the alternating optimization framework produces a non-decreasing and bounded sequence of objective values.
By the monotone convergence principle, such a sequence is guaranteed to converge to a fixed point, corresponding to a locally optimal discrete solution that cannot be further improved by updating either the TX or RX codeword alone.}
\section{Performance Evaluation}
\label{sec:results}
\subsection{Baseline Methods for Comparison}
\label{subsec:initialization} 

\subsubsection{MVDR-CM-HQ}
%The objective function in~(\ref{eq:obj2}) and~(\ref{eq:obj3}) can be maximized using the \Ac{MVDR} beamformer~\cite{10}, when ignoring the \ac{CM}, discrete phase setting and communication constraints. The solution given by the \Ac{MVDR} beamformer is optimal but not CM or discrete, nor does it consider the communication constraint, hence, only provides an engineering upper bound for the problems in~(\ref{eq:obj2}) and~(\ref{eq:obj3}). The analytic solution of the \ac{MVDR} beamformer to the problem in~(\ref{eq:obj2}) is: 
When ignoring the \ac{CM} and discrete phase setting constraints, the objective function in~(\ref{eq:obj2}) can be maximized using the \Ac{MVDR} beamformer~\cite{10}. The solution provided by the \Ac{MVDR} beamformer is optimal but not CM or discrete: 
\begin{align}
	\mathbf{w}_\mathrm{MVDR}(\theta)=\frac{\mathbf{G}^{-1} \mathbf{b}(\theta)}{\mathbf{b}^{H}(\theta) \mathbf{G}^{-1} \mathbf{b}(\theta)},
\end{align}
where the meanings of $\mathbf{G}$ and $\mathbf{b}(\theta)$ are explained in Section~\ref{RX}.
When considering the constraints, the solution found by the MVDR method can be forced to be \ac{CM}, whose phases can be further hard-quantized to the nearest feasible points in the set $e^{j\mathcal{D}}$. We refer to this baseline method as MVDR-CM-HQ. This baseline method serves as a comparison for the proposed FP-SS method for RX codebook optimization with a fixed TX beam. 
\subsubsection{The Effective MVDR beamformer}
The standard MVDR beamformer optimizes objective functions with a single optimization variable and is not applicable to joint TX and RX codebook optimization problems.
In~\cite{888}, the effective MVDR beamformer is derived from the standard MVDR beamformer to maximize the objective function in~(\ref{op}), while disregarding all the constraints:
\begin{align}\mathbf{w}_\mathrm{eff-MVDR}=\frac{\mathbf{R}_\mathrm{eff}^{-1} \mathbf{a}_\mathrm{eff}(\theta)}{\mathbf{a}_\mathrm{eff}^{H}(\theta) \mathbf{R}_\mathrm{eff}^{-1} \mathbf{a}_\mathrm{eff}(\theta)},
\end{align}	
where $\mathbf{a}_{\mathrm{eff}}(\theta)=\text{vec}(\mathbf{a}_{r}(\theta)\mathbf{a}_{t}^{H}(\theta))\in \mathbb{C}^{M\cdot N\times 1}$, $\mathbf{h}_{\mathrm{SI}}= \text{vec}(\mathbf{H}_{\mathrm{SI}})\in \mathbb{C}^{M\cdot N\times 1}$, and $\mathbf{R}_{\mathrm{eff}}=\mathbf{h}_{\mathrm{SI}}\mathbf{h}_{\mathrm{SI}}^{H}+\frac{\sigma_z^{2}}{M\cdot P_{t}}\mathbf{I}_{M\cdot N}$. The objective value given by the effective MVDR beamformer is:
\begin{align}\frac{\left |\ \mathbf{w}_\mathrm{eff-MVDR}^{H} \mathbf{a}_{\mathrm{eff}}(\theta) \right |^{2}}{\mathbf{w}_\mathrm{eff-MVDR}^{H}\mathbf{R}_{\mathrm{eff}}\mathbf{w}_\mathrm{eff-MVDR}}.
\end{align}	
As demonstrated in~\cite{888}, this objective value is only achievable by a pair of $\mathbf{w}_{t}$ and $\mathbf{w}_{r}$ if $\text{mat}(\mathbf{w}_{\text{eff-MVDR}})$ has rank one. Moreover, the effective MVDR beamformer ignores all the constraints in~(\ref{op}), including the CM constraint, phase quantization and the communication requirement. Hence, the effective MVDR beamformer provides an upper bound for the problem~(\ref{op}).

%\subsubsection{CM-GD-SQ}
%Another baseline method is to integrate a differentiable soft quantization function~\cite{16} into the Constant Modulus Gradient Descent (CM-GD) method~\cite{8}. Thereby, the phase quantization loss can be approximated during the gradient descent search. Still, the hard quantization is required at the end.
\subsubsection{Exhaustive Search (ES)} 
\begin{figure}[t]
	\centering
	\subfigure[Scenario $A$.] {\raisebox{-1.65pt}{\includegraphics[height=0.2643\textwidth]{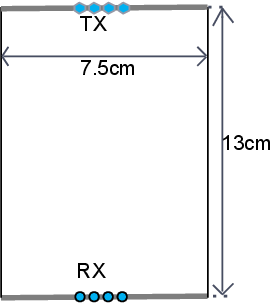}}}
	%\label{fig1a}}
\subfigure[Scenario $B$.] {\includegraphics[height=0.257\textwidth]{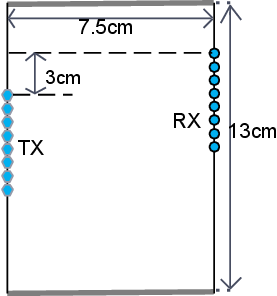}}
%\label{fig1b}}

\caption{Two antenna placement scenarios for performance evaluation.}
\label{fig33}

\end{figure}
In this baseline method, all possible combinations of discrete phase values are exhaustively tested and the combination achieving the highest \ac{SINR} is selected. We denote the ES for RX codebook optimization with a fixed TX codeword as ES-RX, and for TX codebook optimization with a fixed RX codeword as ES-TX. Importantly, ES-TX also considers the communication constraint.
\Ac{ES} is computationally expensive, especially when online codebook synthesis is required.

\subsection{Numerical Results}

To evaluate the proposed methods, we consider two antenna placement scenarios for a mmWave ISAC device, as illustrated in Fig.~\ref{fig33}. 
In Scenario $A$, a TX and an RX ULA, each with $4$ elements, are positioned opposite each other at the center of the device's top and bottom edges, respectively. 
In Scenario $B$, the RX and TX ULAs, each with $8$ elements, are positioned on the left and right edges of the device, with a vertical distance between them. The phase settings are quantized using $8$ bits for Scenario $A$ and $4$ bits for Scenario $B$, respectively. 

We sweep the sensing direction from $-90^{\circ}$ to $90^{\circ}$ with a resolution of $5^{\circ}$. 
\tcc{For each direction, a tentative target is assumed to be located at the worst-case target distance, upon which both the baseline methods and proposed methods are applied to compute the optimal CM discrete codewords.}
For the communication constraint, two possibilities $\theta_c=45^{\circ}$ and $\theta_c=-45^{\circ}$ are considered for both scenarios, with the required value $c=3$ for Scenario $A$ and $c=6$ for Scenario $B$, respectively.
For the evaluation of the proposed discrete RX codebook design method presented in Section~\ref{RX}, the TX codeword $\mathbf{v}_0$ is fixed to be the DFT codeword pointing at the communication direction $\theta_c$; for the evaluation of the proposed discrete TX codebook design method presented in Section~\ref{TX}, the RX codeword $\mathbf{w}_0$ is fixed to be the quantized version of the steering vector pointing to the respective sensing direction.
The transmit power and mmWave carrier frequency are set to $20$ dBm and $28$ GHz, respectively.
For the antenna coupling coefficient $\beta_{n,m}$, the values $G_{2}$ = $0.16$ and $G_{3}$ = $0.67$ are considered, which corresponds to an antenna element on a substrate with $2.2$ in relative permittivity~\cite{14}. 
The TX antenna gain $G_{t,\theta}$ and the RX antenna gain $G_{r,\theta}$ are set to $1$ for all the sensing directions.
\tcc{Based on the work in~\cite{16AA,16BB}, the worst-case RCS is set to $-10$ dBsm.
The worst-case distance is chosen as $10$ m, since in the intended user-end ISAC scenario, the sensing device only needs to detect nearby targets within a range of up to $10$ m.
These parameter selections ensure that the derived $\alpha_{\theta}$ corresponds to the most challenging sensing condition in the considered scenario.}
The noise power is set to $-110$ dBm. 
\begin{figure}[t]
	\centering
	%\vspace{-6mm}
	\subfigure[$\theta_c=45^{\circ}$] {\includegraphics[trim={0.2cm 0.2cm 0.2cm 0.2cm},clip,width=0.48\textwidth]{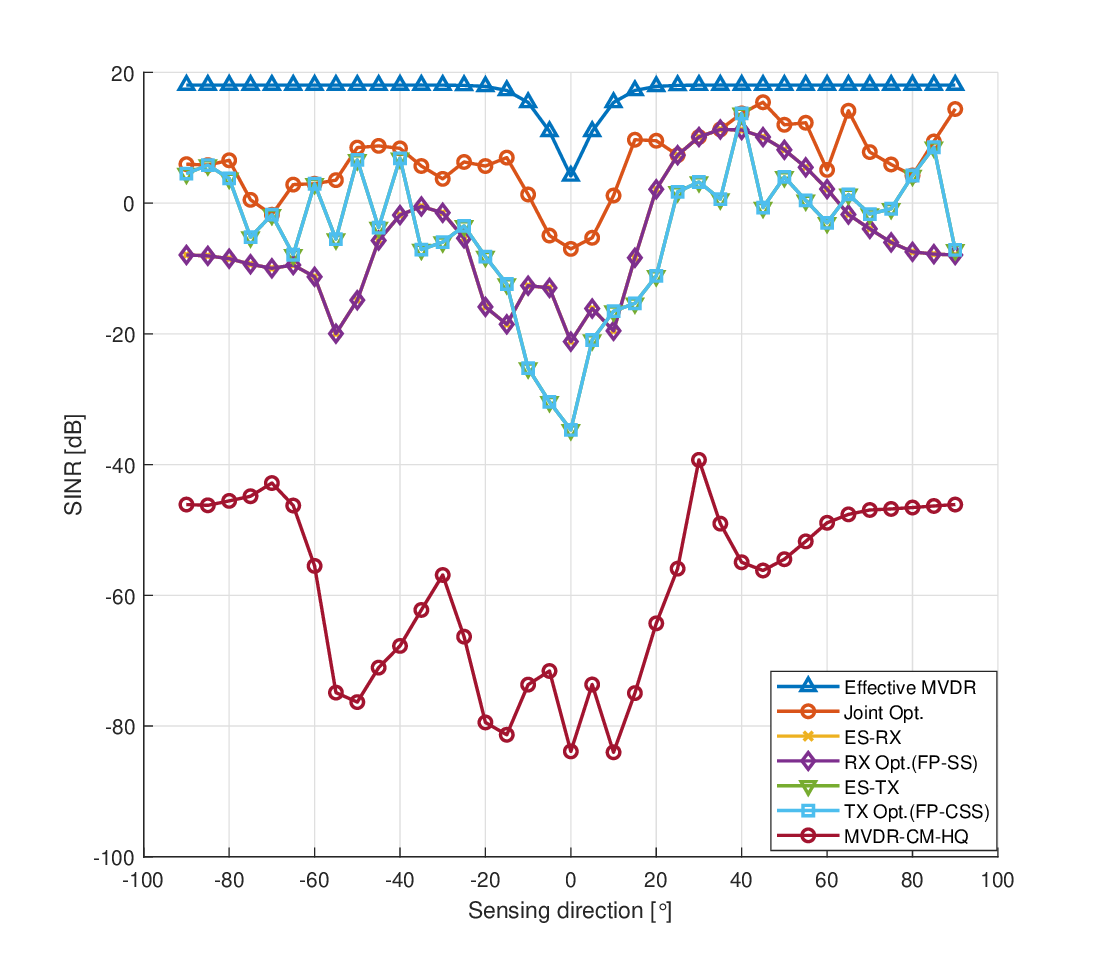}
		\label{fig1a}}
	\subfigure[$\theta_c=-45^{\circ}$] {\includegraphics[trim={0.2cm 0.2cm 0.2cm 0.2cm},clip,width=0.48\textwidth]{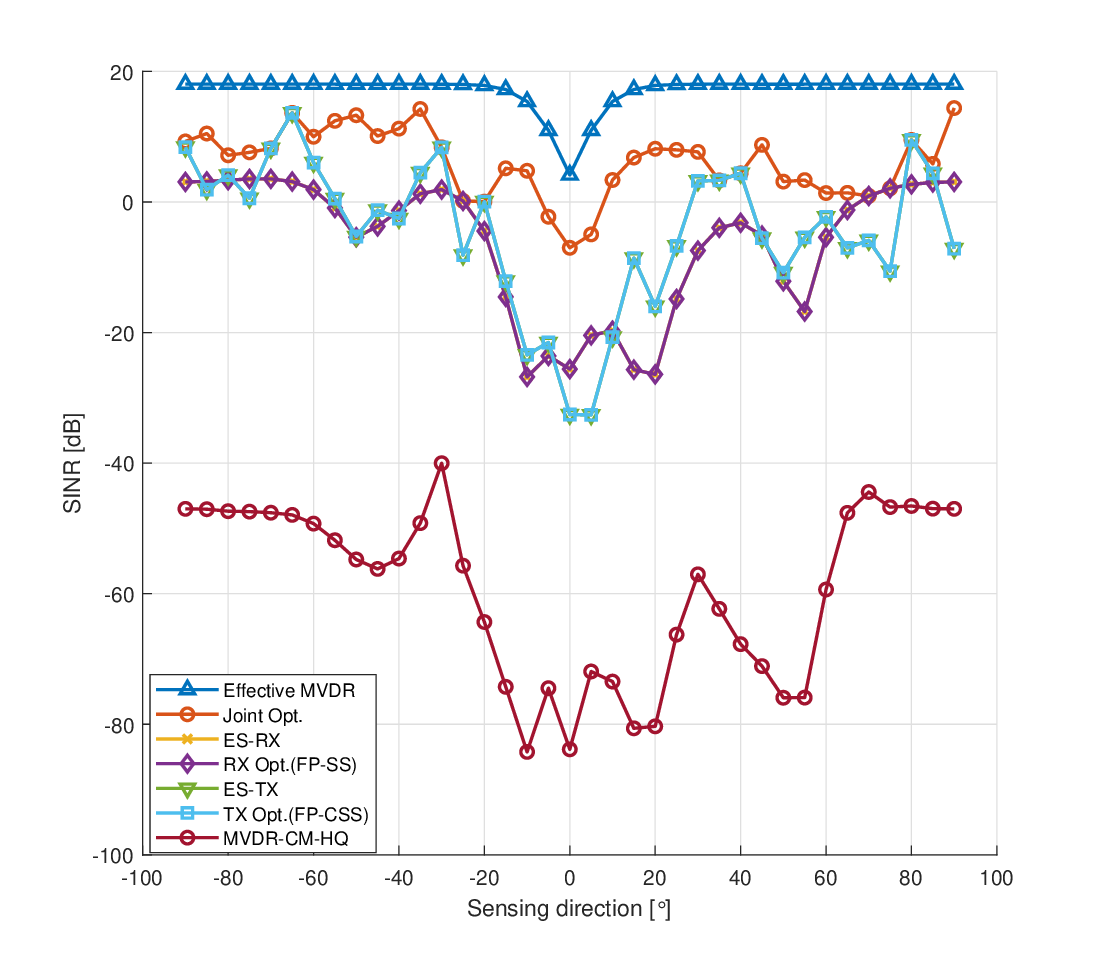}%trim={left down right above}
		\label{fig1b}}
	%\vspace{-3mm}
	\caption{Achieved SINRs by different methods for Scenario $A$ with a communication requirement of $c=3$. (a) corresponds to a communication direction of $\theta_c=45^{\circ}$, while (b) corresponds to $\theta_c=-45^{\circ}$.}
	%\vspace{-1mm}
	\label{fig1}
\end{figure}
\begin{figure}[t]
	\centering
	%\vspace{-6mm}
	\subfigure[$\theta_c=45^{\circ}$] {\includegraphics[trim={0.2cm 0.2cm 0.2cm 0.2cm},clip,width=0.48\textwidth]{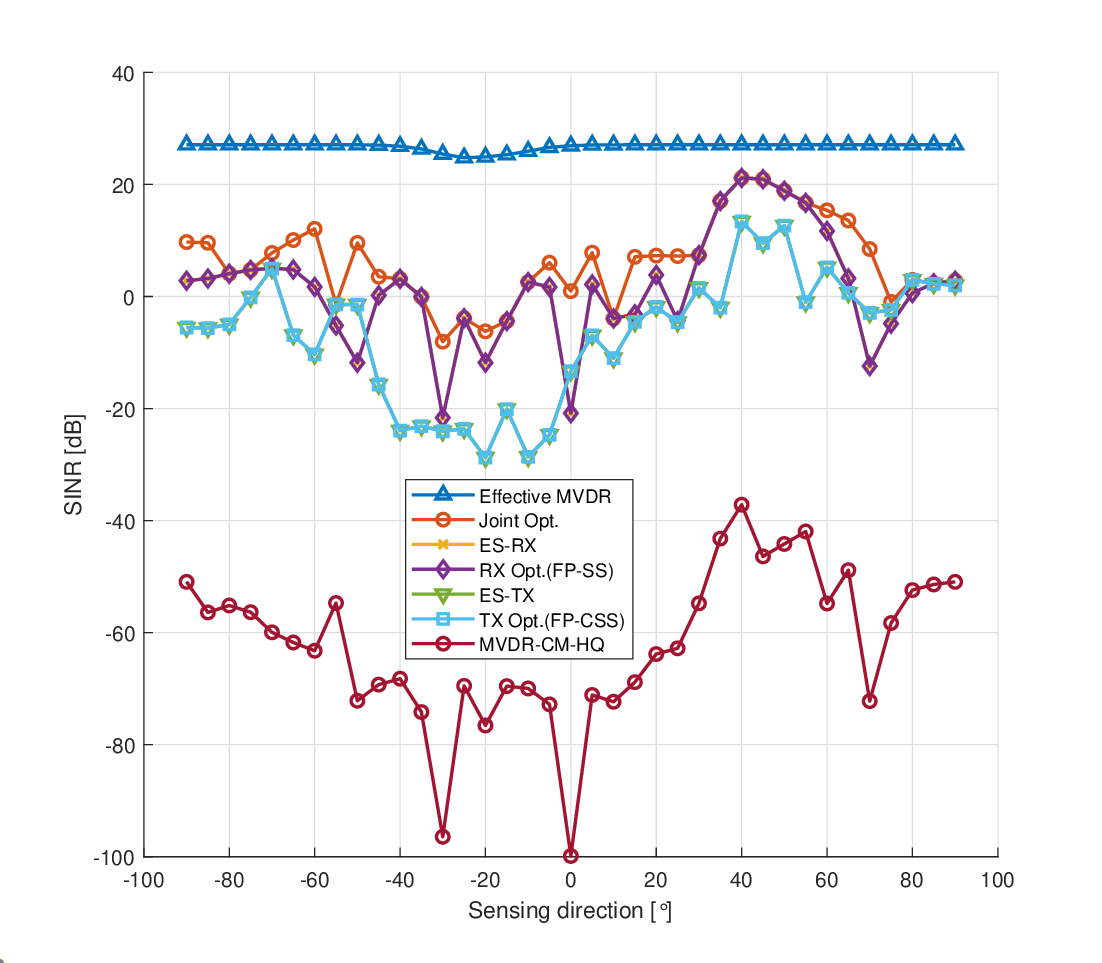}
		\label{fig2a}}
	\subfigure[$\theta_c=-45^{\circ}$] {\includegraphics[trim={0.2cm 0.2cm 0.2cm 0.2cm},clip,width=0.48\textwidth]{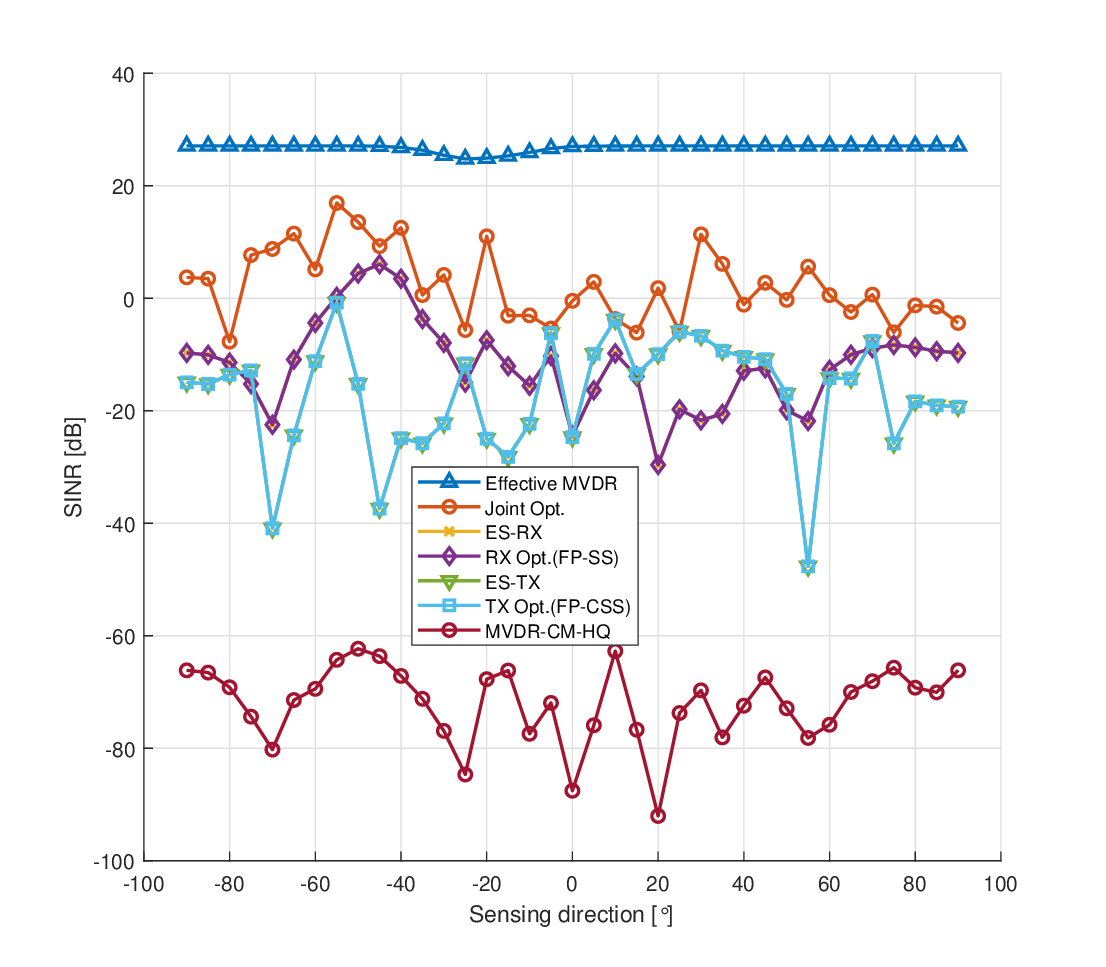}%trim={left down right above}
		\label{fig2b}}
	%\vspace{-3mm}
	\caption{Achieved SINRs by different methods for Scenario $B$ with a communication requirement of $c=6$. (a) corresponds to a communication direction of $\theta_c=45^{\circ}$, while (b) corresponds to $\theta_c=-45^{\circ}$.}
	%\vspace{-1mm}
	\label{fig6}
\end{figure}

  \begin{figure}[t]
	\centering
	%\vspace{-6mm}
	\subfigure[Scenario $A$] {\includegraphics[trim={0.15cm 0.15cm 0.15cm 0.15cm},clip,width=0.48\textwidth]{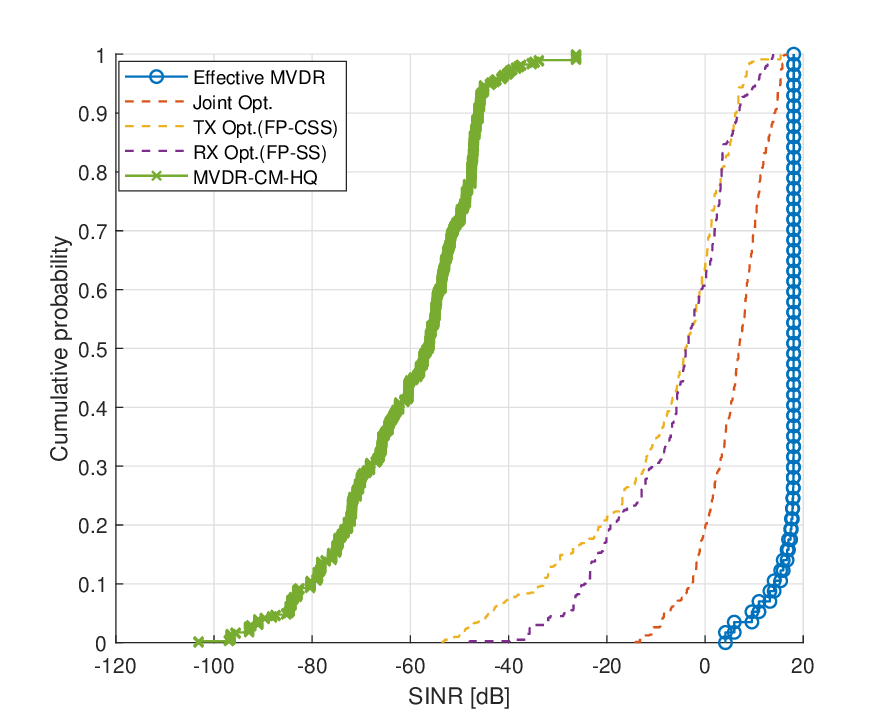}
		\label{fig3a}}
	\subfigure[Scenario $B$] {\includegraphics[trim={0.15cm 0.15cm 0.15cm 0.15cm},clip,width=0.48\textwidth]{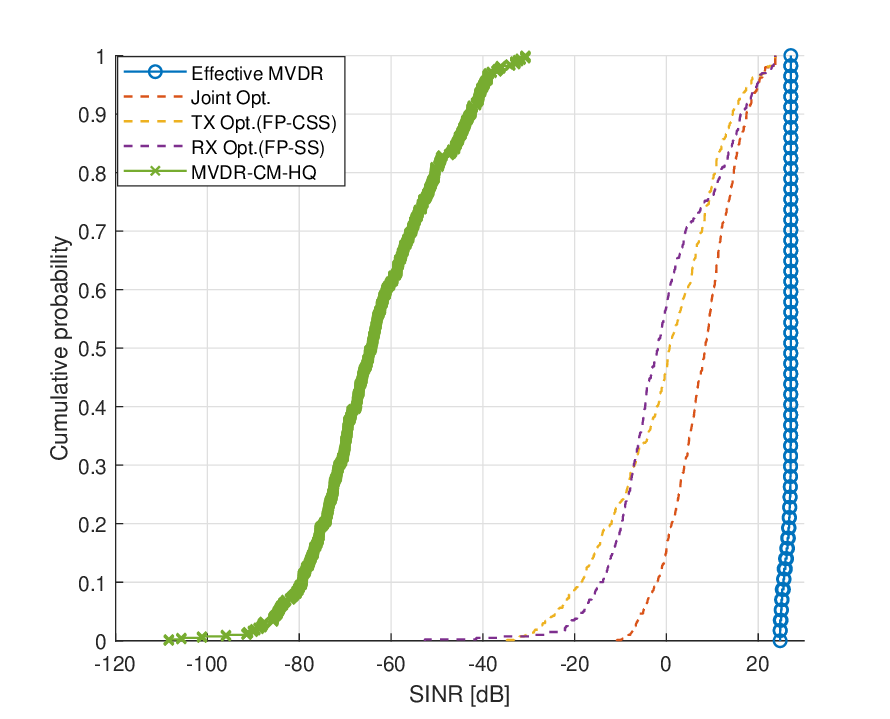}%trim={left down right above}
		\label{fig3b}}
	%\vspace{-3mm}
	\caption{CDF for various sensing directions, with various communication directions $\theta_c$ and varying values of $c$.}
	%\vspace{-1mm}
	\label{fig333}
\end{figure}
 
\begin{figure}[t]
	\centering
	\includegraphics[trim={0.7cm 0cm 0cm 0cm},clip,width=0.36\textwidth]{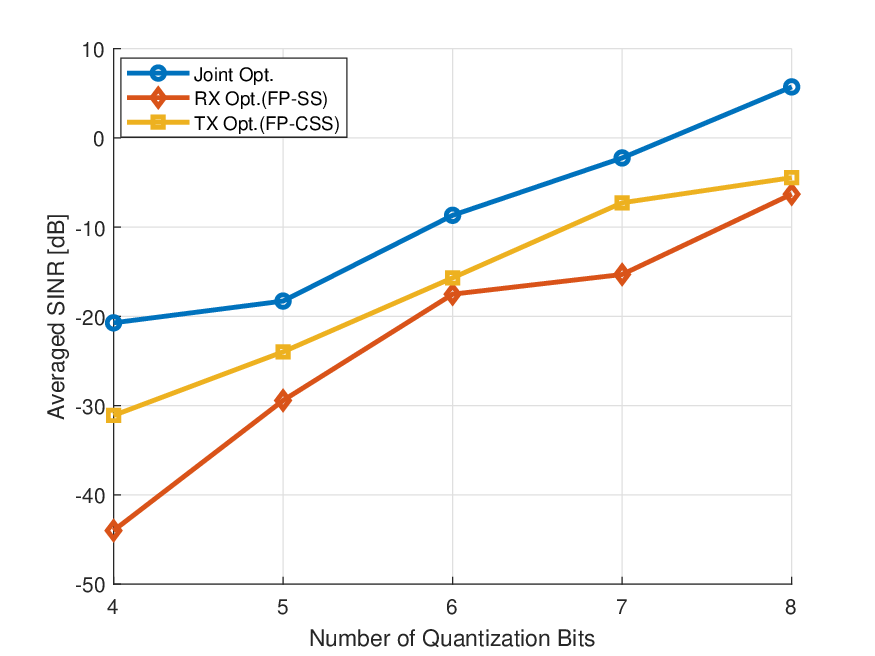}
	\caption{Average sensing SINR versus the number of phase quantization bits for the proposed codebook design methods for Scenario~$A$ with communication direction $\theta_{c} = -45^\circ$ and communication gain requirement $c = 3$. The average SINR is computed over all scanned sensing directions.}
	\label{Addfig1}
	\vspace{-4mm}
\end{figure}

The achieved SINRs over swept sensing directions using different methods are shown in Figs.~\ref{fig1} and~\ref{fig6}.
First, it can be observed that the effective MVDR method achieves the highest SINRs across all sensing directions. However, this result only serves as an upper bound for the joint TX and RX codebook optimization problem~(\ref{op}), as it disregards the CM constraint, the phase quantization and the communication requirement. In all scenarios, the proposed FP-SS method for RX codebook design with a fixed TX codebook, and the FP-CSS method for TX codebook design with a fixed RX codebook, achieve the same SINRs as those obtained by the corresponding ES across all sensing directions, which is optimal for the respective discrete codebook design. This aligns with the theoretical optimality of the proposed FP-SS and FP-CSS methods, as demonstrated in Section~\ref{RX} and~\ref{TX}. In comparison to the proposed FP-SS method for RX codebook design with a fixed TX codeword, the baseline method MVDR-CM-HQ shows significantly worse performance. The proposed joint codebook design method outperforms the RX or TX codebook design method, as it optimizes both TX and RX codewords jointly, rather than fixing one of them. It is also notable that the performance of the proposed joint optimization is relatively close to the effective MVDR beamformer, which serves as the upper bound. \tcc{Note that the effective MVDR beamformer neither satisfies the phase quantization constraint nor meets the communication requirement, while the proposed joint optimization operates under $8$-bit phase quantization and satisfies the communication requirement. Furthermore, when considering a $10$-bit ADC with a required minimal SINR to avoid ADC saturation at $-54$ dB~\cite{FF}, the proposed methods consistently meet this requirement across all sensing directions.}
%The joint codebook design method can improve at some points and at some points not, which is because it already converges at the points where it does not improve anymore.  
It can be further observed that, the achieved SINRs given by the upper bound at $0^{\circ}$ in Scenario $A$ and at $-25^{\circ}$ in Scenario $B$, are worse than those at other sensing directions. This is because the SI channel is more correlated at these two sensing directions, which makes SI more difficult to suppress. Despite that, the FP-SS and FP-CSS methods are able to find the respective optimal discrete codewords. 
 
To evaluate the performance of the different methods from a statistical perspective, the \ac{CDF} of the achieved SINRs for different methods is also plotted. Fig.~\ref{fig333} illustrates the \ac{CDF} of the achieved SINRs across various directions with varying communication directions and specified communication beam gains. 
The evaluated sensing directions range from $-85^{\circ}$ to $85^{\circ}$ with a resolution of $3^{\circ}$, while the communication directions are $\{30^{\circ},40^{\circ},50^{\circ},60^{\circ},70^{\circ},80^{\circ},90^{\circ}\}$. The specified values for communication beam gain, $c$, are set to $2$ and $3$ for Scenario $A$, and $4$ and $6$ for Scenario $B$. 
From Fig.~\ref{fig333}, it can be observed that, from the statistical perspective, the proposed FP-SS method for RX codebook design with a fixed TX beam significantly outperforms the baseline MVDR-CM-HQ method. The proposed FP-CSS method for TX codebook design with a fixed RX beam shows similar performance to the FP-SS method. Furthermore, the joint codebook design method delivers even better performance, compared to the individual RX or TX codebook design method, coming relatively close to the effective MVDR upper bound. These findings are in line with the results shown in Figs.~\ref{fig1} and~\ref{fig6}. The corresponding ES methods achieve the same performance as the proposed FP-SS/FP-CSS method and are therefore not plotted in Fig.~\ref{fig333}.

\subsection{\textcolor{red}{Robustness to AME and Composite SI Channels}}
\label{sec:robustness}

\textcolor{red}{We further evaluate the robustness of the proposed methods under two practical non-idealities: array manifold errors (AMEs) in the far-field sensing response and reflected components in the SI channel. Motivated by mmWave array-calibration studies on element-dependent gain/phase errors~\cite{JJ,KK}, we model AME by perturbing the far-field steering responses as:}
\[
\widetilde{\mathbf a}_r(\theta)=\mathbf D_r \mathbf a_r(\theta), \quad
\widetilde{\mathbf a}_t(\theta)=\mathbf D_t \mathbf a_t(\theta),
\]
\textcolor{red}{where the diagonal entries of the diagonal matrices \(\mathbf D_r\) and \(\mathbf D_t\) are \((1+\epsilon_g)e^{j\epsilon_\phi}\). The perturbations \(\epsilon_g \sim \mathcal{N}(0,\sigma_g^2)\) and \(\epsilon_\phi \sim \mathcal{N}(0,\sigma_\phi^2)\) are independently generated across antenna elements, with \(\sigma_g=0.1\) and \(\sigma_\phi=10^\circ\). The reported SINRs are averaged over \(30\) independent AME realizations. The error bars indicate one standard deviation over these AME realizations.}

\textcolor{red}{The composite SI channel is modeled as:}
\[
\mathbf{H}_{\mathrm{SI}}^{\mathrm{comp}}
=
\mathbf{H}_{\mathrm{SI}}
+
\sum_{\ell=1}^{L_{\mathrm{NLoS}}}
\eta_\ell e^{j\xi_\ell}
\frac{\mathbf a_r(\theta_\ell^R)\mathbf a_t^H(\theta_\ell^T)}
{\left\|\mathbf a_r(\theta_\ell^R)\mathbf a_t^H(\theta_\ell^T)\right\|_F},
\]
\textcolor{red}{where \(\mathbf{H}_{\mathrm{SI}}\) is the near-field coupling channel, $L_{\mathrm{NLoS}}$ is the number of NLoS SI paths, and \(\xi_\ell\) is a random phase uniformly distributed in \([0,2\pi)\). The relative power of the \(\ell\)-th reflected path is defined as}
$
\frac{\eta_\ell^2}{\|\mathbf{H}_{\mathrm{SI}}\|_F^2},
$
\textcolor{red}{where \(\|\cdot\|_F\) denotes the Frobenius norm. 
In Scenario \(A\), the two reflected NLoS paths are \((\theta_\ell^R,\theta_\ell^T)=(0^\circ,0^\circ)\) and \((65^\circ,70^\circ)\), with relative powers \(-30\) dB and \(-25\) dB, respectively. In Scenario \(B\), the corresponding paths are \((30^\circ,35^\circ)\) and \((-65^\circ,-70^\circ)\), with relative powers \(-30\) dB and \(-25\) dB, respectively.}

\begin{figure}[H]
	\centering
	\subfigure[Scenario $A$, $\theta_c=-45^\circ$, $c=3$.] {\includegraphics[width=0.85\columnwidth]{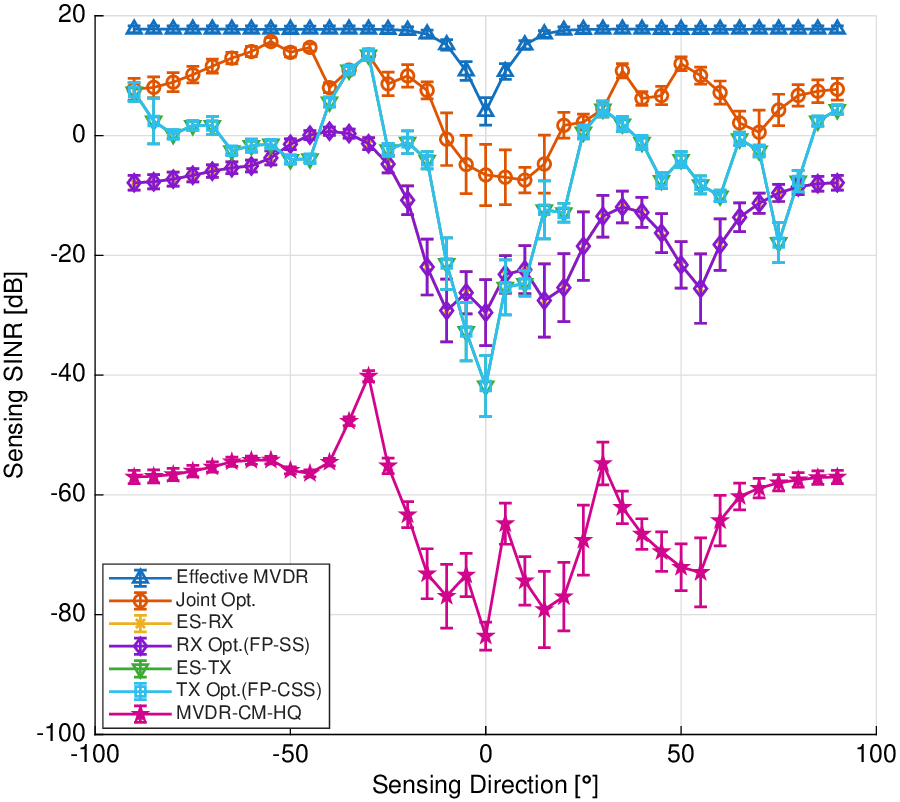}}\\[-0.5mm]
	\subfigure[Scenario $B$, $\theta_c=45^\circ$, $c=6$.] {\includegraphics[width=0.85\columnwidth]{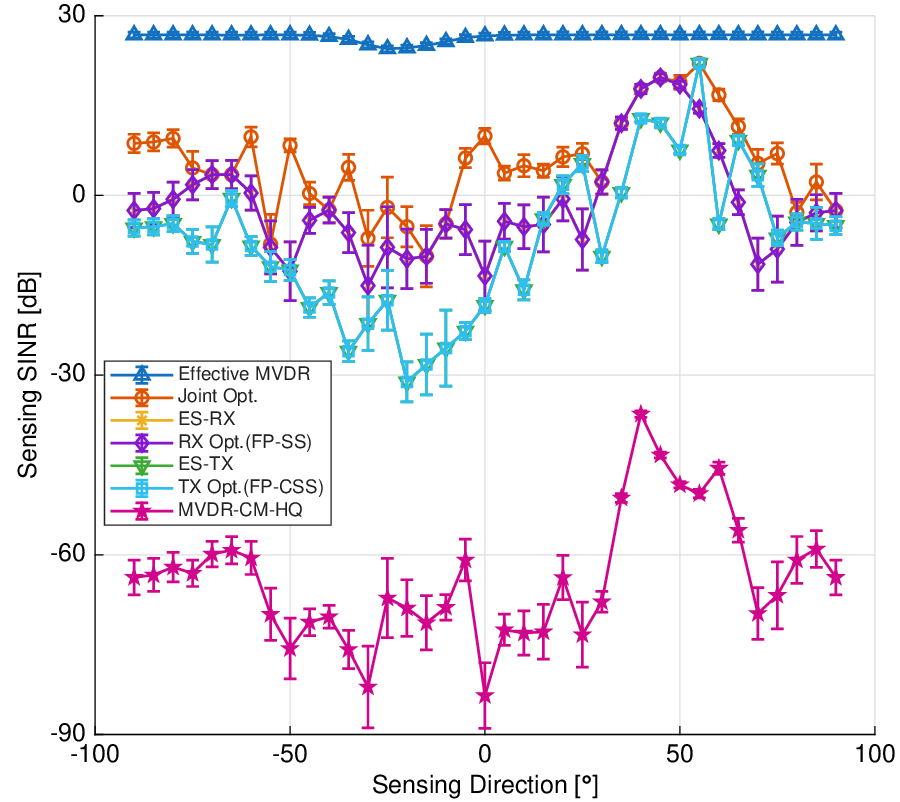}}
	\caption{Robustness evaluation under AME with \(\sigma_g=0.1\) and \(\sigma_\phi=10^\circ\), and composite SI channels with two reflected NLoS paths.}
	\label{fig:robustness}
\end{figure}

\textcolor{red}{The simulation results are shown in Fig.~\ref{fig:robustness}. From the figure we observe that the achieved sensing SINR does not exhibit an abrupt performance collapse under the considered AME level.  Furthermore, under the composite SI channel with reflected NLoS components, the FP-SS and FP-CSS curves coincide with their corresponding ES-RX and ES-TX baselines, respectively. This confirms that the optimality of the proposed discrete search methods holds for a general SI channel matrix \(\mathbf{H}_{\mathrm{SI}}\), not only for the nominal near-field coupling model.}

\tcc{To evaluate the impact of phase quantization on the proposed discrete codebook design framework, we examine the sensing performance under different quantization bit resolutions for Scenario~$A$, with the communication direction fixed at $\theta_c = -45^{\circ}$ and the required communication beamforming gain set to $c = 3$. We vary the number of quantization bits from $4$ to $8$ and apply three optimization strategies: (i) the proposed joint TX-RX codebook design (Joint Opt.), (ii) the RX codebook design based on FP-SS, and (iii) the TX codebook design based on FP-CSS. For each quantization resolution, the sensing SINR is computed over all scanned sensing directions, and the final performance metric is obtained by averaging the SINR values over all sensing directions.
Fig.~\ref{Addfig1} presents the resulting average sensing SINR as a function of the quantization bit resolution. As expected, the sensing performance consistently improves as the number of quantization bits increases for all three methods. In particular, when the quantization resolution increases from $4$ to $8$ bits, all three methods exhibit a clear and consistent improvement in average sensing SINR, with the proposed joint TX-RX optimization achieving the highest performance across all resolutions. These results verify that increasing the phase resolution effectively mitigates quantization loss.}

\begin{figure}[t]
	\centering
	\includegraphics[trim={0cm 0cm 0cm 0cm},clip,width=0.36\textwidth]{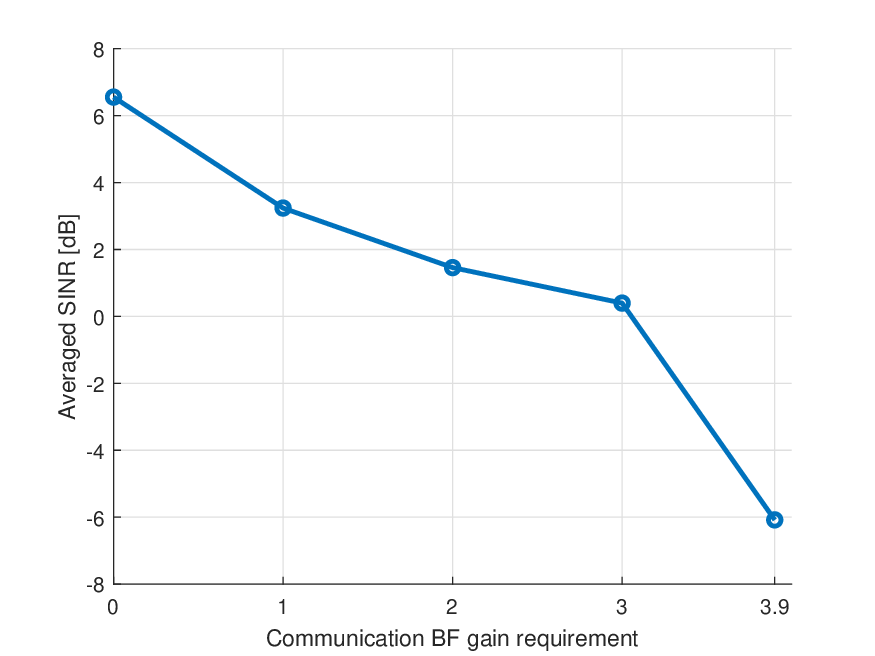}
	\caption{Trade-off between sensing performance and communication beamforming gain requirement $c$ in Scenario~$A$ (communication direction $\theta_{c} = 0^\circ$) using the proposed joint TX-RX codebook design method.
	}
	\label{Addfig2}
	\vspace{-4mm}
\end{figure}

\tcc{To illustrate the trade-off between sensing performance and communication beamforming gain requirement, we further investigate the impact of the communication constraint parameter $c$ in Scenario $A$, with the communication direction fixed at  $\theta_{c} = 0^\circ$. In this experiment, we fix the phase quantization resolution to $8$ bits and apply the proposed joint TX-RX codebook design method for different values of $c$. Specifically, we consider $c \in \{0, 1, 2, 3, 3.9\}$, which spans from the case with no communication constraint ($c = 0$) to a very stringent communication gain requirement. 
The value $c = 3.9$ is chosen close to the maximum achievable beamforming gain of the discrete codebook at $\theta_{c}$; for $c = 4$, the optimization problem becomes infeasible due to phase quantization. The resulting trade-off is shown in Fig.~\ref{Addfig2}, where the average sensing SINR over the scanned sensing directions decreases as the 
communication gain requirement $c$ increases.}

\begin{figure}[t]
 	\centering
 	\includegraphics[trim={0cm 0cm 0cm 0cm},clip,width=0.36\textwidth]{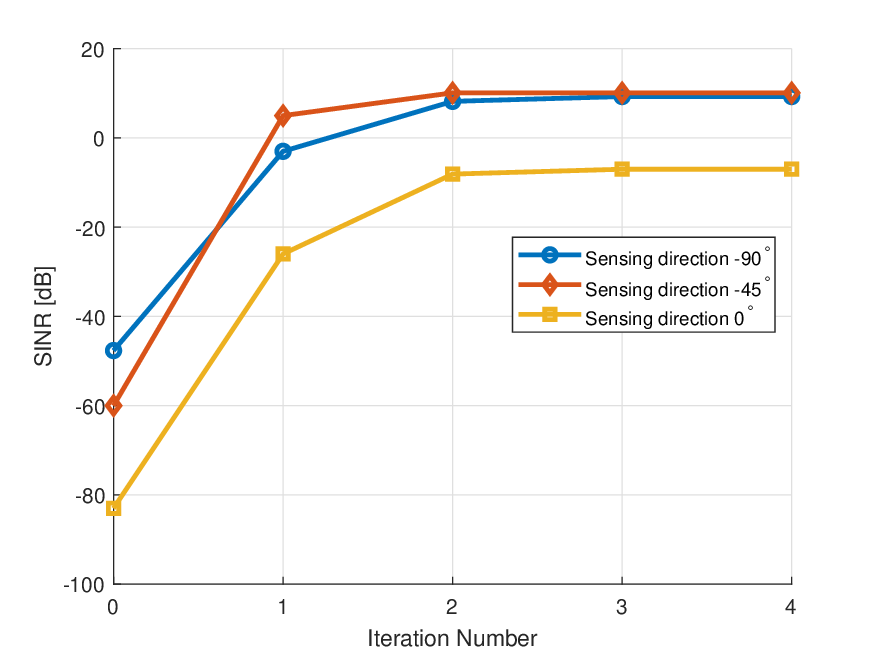}
 	\caption{Convergence behavior of the proposed joint codebook optimization algorithm for three representative sensing directions under Scenario~$A$ with communication direction \(\theta_c=-45^\circ\), \(c=3\), and $8$-bit phase quantization.}
 	\label{Addfig3}
 	\vspace{-2mm}
 \end{figure}

\tcc{To complement the theoretical convergence discussion in Section~\ref{joint}, Fig.~\ref{Addfig3} presents the evolution of the sensing SINR across iterations for three representative sensing directions, namely \(-90^\circ\), \(-45^\circ\), and \(0^\circ\), under Scenario~$A$ with the communication requirement $\theta_c = -45^\circ$, \(c=3\) and an 8-bit phase quantization resolution.  
As shown in the figure, the sensing SINR improves significantly from iteration $0$ to iteration $1$ and then gradually converges within fewer than five iterations for all tested directions.  
This behavior verifies the convergence property analyzed in Section~\ref{joint}.}
 
To evaluate the complexity of the proposed methods, the numbers of floating-point operations (FLOPs)~\cite{17} are measured for the proposed joint codebook design method, \mbox{FP-SS}, and FP-CSS method in both scenarios. 
\textcolor{red}{To enable a consistent comparison across different optimization schemes, the measured FLOPs are normalized w.r.t. the ES-joint-TX-RX, i.e., the exhaustive joint search over all feasible TX-RX codeword pairs. Table~\ref{table} reports the resulting normalized complexity for both scenarios. 
	%The full ES-joint-TX-RX SINR sweep is computationally prohibitive, requiring on average $6.7331\times10^{14}$ and $1.8383\times10^{16}$ FLOPs per sensing direction in Scenarios~$A$ and $B$, respectively. 
	In both scenarios $A$ and $B$, the FP-SS and FP-CSS methods incur only a small fraction of the complexity of their corresponding exhaustive search baselines, ES-RX and ES-TX, respectively, while achieving the same SINR performance.}
\begin{table}[t]
	\centering
	\caption{\revmark{Normalized Number of FLOPs w.r.t. ES-joint-TX-RX}}
	\begin{tabular}{|c|c|c|}
		\hline
		\diagbox{Algorithms}{Scenarios}&Scenario $A$&Scenario $B$\\ 
		\hline
		Joint Opt.&\revmark{$1.5093\times 10^{-8}$}&\revmark{$5.2556\times 10^{-8}$}\\
		\hline
		FP-SS&\revmark{$2.2798\times 10^{-9}$}&\revmark{$2.8875\times 10^{-8}$}\\
		\hline
		FP-CSS&\revmark{$1.0936\times 10^{-8}$}&\revmark{$1.4944\times 10^{-8}$}\\
		\hline
		\revmark{ES-RX}&\revmark{$1.8688\times 10^{-6}$}&\revmark{$4.0740\times 10^{-6}$}\\
		\hline
		\revmark{ES-TX}&\revmark{$4.9408\times 10^{-7}$}&\revmark{$3.7711\times 10^{-7}$}\\
		\hline
	\end{tabular}
	\label{table}
	\vspace{-5mm}
\end{table}
\section{Conclusions}
\label{sec:conclusion}
In this paper, we exploit RX and TX analog phased arrays with discrete phase settings to suppress the \ac{SI} for \ac{mmWave} \ac{ISAC} devices. We formulated a receiver \ac{SINR} maximization problem that optimizes both RX and TX codewords, aiming to suppress the near-field \ac{SI} signal while maintaining the beamforming gain in the far-field sensing directions. The formulation considers practical constraints, including discrete RX and TX codebooks with quantized phase settings and a communication requirement. To solve the formulated problem, we first propose the FP-SS method for RX discrete codebook design with a fixed TX beam, and then propose the FP-CSS method for TX discrete codebook design with a fixed RX beam. Afterwards, we combine these two methods into a joint TX-RX codebook design approach, which alternately optimizes the RX and TX codewords. Simulation results demonstrate the effectiveness and low computational complexity of the proposed methods. \textcolor{red}{Additional robustness simulations confirm that the
sensing-SINR performance remains effective under AME
and composite SI channels with reflected NLoS components.} Further research work could focus on extending the proposed methods to support other types of antenna arrays, such as uniform rectangular arrays or sparse arrays.
	
{\appendices
	\section{Proof of Optimality and Convergence}\label{appendix1}
	
	In this appendix, we provide the proof for the propositions on optimality and convergence in Subsection~\ref{optimality}. 
	%The proof is provided here as a self-contained justification that each FP subproblem under the discrete phase constraint admits a globally optimal solution via the proposed SS/CSS algorithm; it is not presented as the main theoretical contribution of this paper.
	%\tcc{The proof follows the structure of the continuous-domain argument in~\cite{18}, but is adapted to the discrete feasible set imposed by the phase quantization constraint.}
	
		Let $f(\mathbf{w})=\mathbf{w}^{H}\mathbf{B}(\theta)\mathbf{w}$, $g(\mathbf{w})=\mathbf{w}^{H}\mathbf{G}\mathbf{w}$, the feasible region be $\mathcal{S}$: $\{\mathbf{w} \mid \mathbf{w}=[e^{j \varphi_1 },...,e^{j \varphi_{N-1}},1 ] ^T,  \text{with}\  \varphi_1,...,\varphi_{N-1} \in \mathcal{D}\}$, $F(\rho)=\underset{\mathbf{w}}{\max}  \{f(\mathbf{w})-\rho g(\mathbf{w})\mid\mathbf{w}\in \mathcal{S}\}$, and $Q(\mathbf{w})=f(\mathbf{w})/g(\mathbf{w})$. It holds that $f(\mathbf{w})\geq0$ and $g(\mathbf{w})>0$ for any $\mathbf{w}\in \mathcal{S}$, since $\mathbf{B}(\theta)$ is \ac{PSD} and $\mathbf{G}$ is \ac{PD}.
	First consider the following lemmas, before proving Proposition $1$.
	
	%\textbf{\textit{Lemma $1$}}: The function $F(\rho)=\underset{\mathbf{w}}{\max}  \{f(\mathbf{w})-\rho g(\mathbf{w})\mid\mathbf{w}\in \mathcal{S}\}$ is convex.
	
	%\textit{Proof}: Let $\mathbf{w}_o$ be the optimal solution of $F(j\rho'+(1-j)\rho'')$ with $\rho'\ne\rho''$ and $0\leq j\leq 1$. Thus, we have:
%\begin{equation}
%\begin{aligned}
%	 F&\left(j \rho^{\prime}+(1-j) \rho^{\prime \prime}\right) \\
%	&=  f\left(\mathbf{w}_o\right)-\left(j \rho^{\prime}+(1-j) \rho^{\prime \prime}\right) g\left(\mathbf{w}_o\right)\\
%	&= j\left[f\left(\mathbf{w}_o\right)-\rho^{\prime} g\left(\mathbf{w}_o\right)\right]+(1-j)\left[f\left(\mathbf{w}_o\right)-\rho^{\prime \prime} g\left(\mathbf{w}_o\right)\right] \\
%	&\leq  j \cdot \underset{\mathbf{w}}{\max} \left\{f\left(\mathbf{w}\right)-\rho^{\prime} g\left(\mathbf{w}\right) \mid \mathbf{w} \in S\right\}\\
%	&\ \ \ +(1-j) \cdot \underset{\mathbf{w}}{\max} \left\{f\left(\mathbf{w}\right)-\rho^{\prime \prime} g\left(\mathbf{w}\right) \mid \mathbf{w} \in S\right\} \\
%	&=  j F\left(\rho^{\prime}\right)+(1-j) F\left(\rho^{\prime \prime}\right)\; \square 
%\end{aligned}
%\end{equation}		

\textbf{Lemma 1.} For $\rho' < \rho''$, we have $F(\rho') > F(\rho'')$, i.e., $F(\rho) = \underset{\mathbf{w}}{\max} \{ f(\mathbf{w}) - \rho g(\mathbf{w}) \mid \mathbf{w} \in \mathcal{S} \}$ is strictly monotonic decreasing.

\textbf{Proof.} Let $\mathbf{w}'' \in \mathcal{S}$ be an optimal solution that maximizes $F(\rho'')$. Since $g(\mathbf{w}'') > 0$, we have:
\begin{flalign}
	F(\rho'') &= \underset{\mathbf{w}}{\max} \{ f(\mathbf{w}) - \rho'' g(\mathbf{w}) \mid \mathbf{w} \in \mathcal{S} \} \nonumber\\
	&= f(\mathbf{w}'') - \rho'' g(\mathbf{w}'') \\
	&< f(\mathbf{w}'') - \rho' g(\mathbf{w}'') \nonumber\\
	&\leq \underset{\mathbf{w}}{\max} \{ f(\mathbf{w}) - \rho' g(\mathbf{w}) \mid \mathbf{w} \in \mathcal{S} \} \nonumber
	= F(\rho')  \nonumber& \hfill \square
\end{flalign}

\textbf{Lemma 2.} $F(\rho) = 0$ has a unique solution.

\textbf{Proof.} It is evident that for $g(\mathbf{w}) > 0$, $\underset{\rho \to -\infty}{\lim} F(\rho) = +\infty$ and $\underset{\rho \to +\infty}{\lim} F(\rho) = -\infty$. Furthermore, based on Lemma 1, we can conclude that $F(\rho) = 0$ has a unique solution. \hfill $\square$

\textbf{Lemma 3.} For any $\mathbf{w}' \in \mathcal{S}$, let $\rho' = f(\mathbf{w}') / g(\mathbf{w}')$. Then, we have $F(\rho') \geq 0$.

\textbf{Proof.} $F(\rho') = \underset{\mathbf{w}}{\max} \{f(\mathbf{w}) - \rho' g(\mathbf{w}) \mid \mathbf{w} \in \mathcal{S}\} \geq f(\mathbf{w}') - \rho' g(\mathbf{w}') = 0$. \hfill $\square$ 

\textbf{Proof of Proposition 1.}

\begin{enumerate}
	\item[(a)] Let $\mathbf{w}^* \in \mathcal{S}$ be an optimal solution of $\underset{\mathbf{w}}{\max} \{ f(\mathbf{w}) - \rho^* g(\mathbf{w}) \mid \mathbf{w} \in \mathcal{S} \}$ such that $f(\mathbf{w}^*) - \rho^* g(\mathbf{w}^*) = 0$ and $f(\mathbf{w}) - \rho^* g(\mathbf{w}) \leq f(\mathbf{w}^*) - \rho^* g(\mathbf{w}^*) = 0$, $\forall \mathbf{w} \in \mathcal{S}$.
	
	Since $g(\mathbf{w}) > 0$, we have $\rho^* = f(\mathbf{w}^*) / g(\mathbf{w}^*) \geq f(\mathbf{w}) / g(\mathbf{w})$, $\forall \mathbf{w} \in \mathcal{S}$.
	
	Thus, $\rho^*$ is the maximum of $\underset{\mathbf{w}}{\max} \{ f(\mathbf{w}) / g(\mathbf{w}) \mid \mathbf{w} \in \mathcal{S} \}$ and $\mathbf{w}^*$ is an optimal solution of $\underset{\mathbf{w}}{\max} \{ f(\mathbf{w}) / g(\mathbf{w}) \mid \mathbf{w} \in \mathcal{S} \}$.
	
	\item[(b)] Let $\mathbf{w}^*$ be an optimal solution of $\underset{\mathbf{w}}{\max} \{ f(\mathbf{w}) / g(\mathbf{w}) \mid \mathbf{w} \in \mathcal{S} \}$ and $\rho^*$ be the optimal objective function value, so we have $\rho^* = f(\mathbf{w}^*) / g(\mathbf{w}^*) \geq f(\mathbf{w}) / g(\mathbf{w})$, $\forall \mathbf{w} \in \mathcal{S}$.
	
	Since $g(\mathbf{w}) > 0$, we have $f(\mathbf{w}) - \rho^* g(\mathbf{w}) \leq f(\mathbf{w}^*) - \rho^* g(\mathbf{w}^*) = 0$, $\forall \mathbf{w} \in \mathcal{S}$.
	
	This implies that $\mathbf{w}^*$ is an optimal solution of $F(\rho^*)=\underset{\mathbf{w}}{\max} \{ f(\mathbf{w}) - \rho^* g(\mathbf{w}) \mid \mathbf{w} \in \mathcal{S} \}$, and that $F(\rho^*)=0$. \hfill $\square$
\end{enumerate}

Before proving Proposition 2, consider the following lemmas.

\textbf{Lemma 4.} Let $\mathbf{w}'$ and $\mathbf{w}''$ be optimal solutions of $F(\rho')$ and $F(\rho'')$, respectively. If $\rho' < \rho''$, then $g(\mathbf{w}') \geq g(\mathbf{w}'')$.

\textbf{Proof.} Since $\mathbf{w}'$ and $\mathbf{w}''$ are optimal solutions of $F(\rho')$ and $F(\rho'')$, respectively, we have:
\begin{align}
	f(\mathbf{w}') - \rho' g(\mathbf{w}') \geq f(\mathbf{w}'') - \rho' g(\mathbf{w}'') \\
	f(\mathbf{w}'') - \rho'' g(\mathbf{w}'') \geq f(\mathbf{w}') - \rho'' g(\mathbf{w}')
\end{align}

Adding the above two inequalities and rearranging both sides of the new one, we have:
\begin{align}
	(\rho'' - \rho') g(\mathbf{w}') &\geq (\rho'' - \rho') g(\mathbf{w}'')
\end{align}

As $\rho''-\rho' > 0$, we have $g(\mathbf{w}') \geq g(\mathbf{w}'')$. \hfill $\square$

\textbf{Lemma 5.} Let $\mathbf{w}'$ and $\mathbf{w}''$ be optimal solutions of $F(\rho')$ and $F(\rho'')$, respectively. Then, 
\begin{align}
Q(\mathbf{w}'') - Q(\mathbf{w}') \geq \left( \frac{F(\rho'')}{g(\mathbf{w}'')} \right) - \left( \frac{F(\rho'')}{g(\mathbf{w}')} \right).
\end{align}

\textbf{Proof.} As $\mathbf{w}''$ is an optimal solution of $F(\rho'')$, we have:
\begin{align}
f(\mathbf{w}'') - \rho'' g(\mathbf{w}'') \geq f(\mathbf{w}') - \rho'' g(\mathbf{w}').
\end{align}
Dividing both sides by $g(\mathbf{w}')$, we have:
\begin{align}
\left( \frac{f(\mathbf{w}'')}{g(\mathbf{w}')} \right) - \left( \frac{\rho'' g(\mathbf{w}'')}{g(\mathbf{w}')} \right) \geq \left( \frac{f(\mathbf{w}')}{g(\mathbf{w}')} \right) - \rho''.
\end{align}
Thus, we have:
\begin{align}
	Q&(\mathbf{w}'') - Q(\mathbf{w}') \notag \\
	&= \frac{f(\mathbf{w}'')}{g(\mathbf{w}'')} - \frac{f(\mathbf{w}')}{g(\mathbf{w}')} \\
	&\geq \frac{f(\mathbf{w}'')}{g(\mathbf{w}'')} - \frac{f(\mathbf{w}'')}{g(\mathbf{w}')} + \rho'' \left[ \frac{g(\mathbf{w}'')}{g(\mathbf{w}')} - \frac{g(\mathbf{w}'')}{g(\mathbf{w}'')} \right] \notag  \\
	&= \frac{F(\rho'')}{g(\mathbf{w}'')} - \frac{F(\rho'')}{g(\mathbf{w}')}.&\notag  \hfill \square
\end{align}

\textbf{Lemma 6.} Let $\mathbf{w}'$ and $\mathbf{w}''$ be the optimal solutions of $F(\rho')$ and $F(\rho'')$, respectively. If $\rho' \leq \rho'' \leq \rho^*$, then $Q(\mathbf{w}') \leq Q(\mathbf{w}'')$, where $\rho^*$ satisfies $F(\rho^*) = 0$.

\textbf{Proof.} The proof readily follows from Lemmas 3--5. \hfill $\square$

\textbf{Lemma 7.} Let $\mathbf{w}'$ and $\mathbf{w}''$ be the optimal solutions of $F(\rho')$ and $F(\rho'')$, respectively. Then, 
\begin{align}
Q&(\mathbf{w}'') - Q(\mathbf{w}')\nonumber \\&\leq [-F(\rho'') + (\rho' - \rho'') g(\mathbf{w}'')] \left( \frac{1}{g(\mathbf{w}')} - \frac{1}{g(\mathbf{w}'')} \right).
\end{align}

\textbf{Proof.}  We have $f(\mathbf{w}') - \rho' g(\mathbf{w}') \geq f(\mathbf{w}'') - \rho' g(\mathbf{w}'')$.

 Dividing both sides by $g(\mathbf{w}')$, we have: 
\begin{align}
 \frac{f(\mathbf{w}')}{g(\mathbf{w}')}  - \rho' \geq  \frac{f(\mathbf{w}'')}{g(\mathbf{w}')} - \rho'  \frac{g(\mathbf{w}'')}{g(\mathbf{w}')}  .
\end{align}

Thus, we have:
\begin{align}
	Q&(\mathbf{w}'') - Q(\mathbf{w}')\notag\\
	&= \frac{f(\mathbf{w}'')}{g(\mathbf{w}'')} - \frac{f(\mathbf{w}')}{g(\mathbf{w}')} \notag \\
	&\leq \frac{f(\mathbf{w}'')}{g(\mathbf{w}'')} -  \frac{f(\mathbf{w}'')}{g(\mathbf{w}')} + \rho' \left( \frac{g(\mathbf{w}'')}{g(\mathbf{w}')} - \frac{g(\mathbf{w}'')}{g(\mathbf{w}'')} \right)   \\
	&= \left(-f(\mathbf{w}'') + \rho' g(\mathbf{w}'')\right) \left( \frac{1}{g(\mathbf{w}')} - \frac{1}{g(\mathbf{w}'')} \right) \notag \\
	&= \left( -F(\rho'') + (\rho' - \rho'') g(\mathbf{w}'') \right) \left( \frac{1}{g(\mathbf{w}')} - \frac{1}{g(\mathbf{w}'')} \right).\notag &\hfill \square
\end{align}

\textbf{Lemma 8.} Let $\mathbf{w}'$ and $\mathbf{w}^*$ be the optimal solutions of $F(\rho')$ and $F(\rho^*)$, where $\rho^*$ satisfies $F(\rho^*) = 0$. Then we have:
\begin{align}
\rho^* - Q(\mathbf{w}') \leq (\rho^* - \rho') \left( 1 - \frac{g(\mathbf{w}^*)}{g(\mathbf{w}')} \right).
\end{align}

\textbf{Proof.} Since $\rho^*$ satisfies $F(\rho^*) = 0$, we have $Q(\mathbf{w}^*) = \rho^*$. Based on Lemma 7, it is easy to prove Lemma 8. \hfill $\square$

\textbf{Proof of Proposition 2.} Dinkelbach's algorithm updates $\rho_i$ with the previous value of $Q(\mathbf{w})$, i.e., $\rho_{i+1} = Q(\mathbf{w}_i)$, where $\mathbf{w}_i$ is the optimal solution of $F(\rho_i)$. From Lemma 8, the algorithm converges with a rate of $\left( 1 - \frac{g(\mathbf{w}^*)}{g(\mathbf{w}_i)} \right)$. From Lemma~$4$, we have that $\left( 1 - \frac{g(\mathbf{w}^*)}{g(\mathbf{w}_i)} \right)$ is nonincreasing. Therefore, the sequence $\{ \rho_i \}$ obtained by Dinkelbach's algorithm converges superlinearly to $\rho^*$ for each $\rho_i < \rho^*$. \hfill $\square$

	\section{Intersection of two allowable ranges}\label{appendix2}
	Assume that $\tilde{\eta}_1$ and $\tilde{\eta}_2$ are calculated based on (\ref{49}) and (\ref{53}) for antenna element $m$, respectively. In this appendix, we describe how to determine the final allowable range for $\psi_{m}$ that does not violate the inequality~(\ref{constraint}) or (\ref{eq:13}). 
	
	If $\tilde{\eta}_1>1$ or $\tilde{\eta}_2>1,$ the final allowable range for $\psi_{m}$ is empty. Otherwise, we need to determine two separate ranges based on $\tilde{\eta}_1$ and $\tilde{\eta}_2$, and then calculate their intersection to determine the final allowable values for $\psi_{m}$. 
	
	If $\tilde{\eta}_1<-1$, the preliminary allowable range obtained based on $\tilde{\eta}_1$ is $[0,2\pi]$. If $-1\leq\tilde{\eta}_1\leq 1$, the range obtained based on $\tilde{\eta}_1$ is: 
	\begin{align}\left[\left\lceil\left(\widehat{\psi}_{m}-\cos ^{-1} \tilde{\eta}_1\right)\right\rceil ,\left\lfloor\left(\widehat{\psi}_{m}+\cos ^{-1} \tilde{\eta}_1\right)\right\rfloor\right].
	\end{align} 
	Similarly, if $\tilde{\eta}_2<-1$, the preliminary allowable range obtained based on $\tilde{\eta}_2$ is $[0,2\pi]$. If $-1\leq\tilde{\eta}_2\leq 1$, the range obtained based on $\tilde{\eta}_2$ is: 
	\begin{align}\left[\left\lceil\left(\widehat{\psi}_{m}'-\cos ^{-1} \tilde{\eta}_2\right)\right\rceil ,\left\lfloor\left(\widehat{\psi}_{m}'+\cos ^{-1} \tilde{\eta}_2\right)\right\rfloor\right].
	\end{align}
	 For the case of $\tilde{\eta}_1\leq1$ and $\tilde{\eta}_2\leq1$, we denote the two preliminary allowable ranges as $[l_1,u_1]$ and $[l_2,u_2]$. We have $l_1,l_2\in[-\pi,2\pi]$, $u_1,u_2\in[0,3\pi]$. Furthermore, it holds that $u_1-l_1\leq 2\pi$ and $u_2-l_2\leq 2\pi$. Below, we present how to get the intersection of these two ranges.
	
	 \noindent \textbf{STEP 1. Normalize the ranges.} 
	
	 \noindent The normalization criterion is that the lower bounds $l_1$ and $l_2$ should lie within the range $[0,2\pi]$. For example, if $l_1 \in [-\pi,0)$, the normalized range will be $[l_1+2\pi,u_1+2\pi]$.
	If $l_1 \in [0,2\pi]$, the range remains unchanged as $[l_1,u_1]$. Denote the two normalized ranges as $[l'_1,u'_1]$ and $[l'_2,u'_2]$. After normalization, $l'_1,l'_2\in[0,2\pi]$ and $u'_1,u'_2\in[0,4\pi]$. 
	
	 \noindent\textbf{STEP 2. Find the intersection.} 
	
	\begin{itemize}
		\item If both upper bounds, $u_1'$ and $u_2'$, lie within the range $[0,2\pi]$, the intersection is $[\max(l_1',l_2'), \min(u_1',u_2')]$;
		\item If only one of the upper bounds lies within $[0,2\pi]$ and the other lies within $[2\pi, 4\pi]$, assume that $ u_1'\in[0,2\pi]$ and $ u_2'\in[2\pi,4\pi]$. In this case, the 
		intersection is $[\max(l_1',l_2'), u_1']\cup [l_1'+2\pi, \min(u_1'+2\pi, u_2')]$;
		\item If both upper bounds, $ u_1'$ and $ u_2'$, lie within $[2\pi,4\pi]$, assume that $ u_1'\leq u_2'$. The intersection is $[\max(l_1',l_2'), u_1']\cup [l_1'+2\pi, u_2'].$
	\end{itemize}
	}

\begin{IEEEbiography}[{\includegraphics[width=1in,height=1.25in,clip,keepaspectratio]{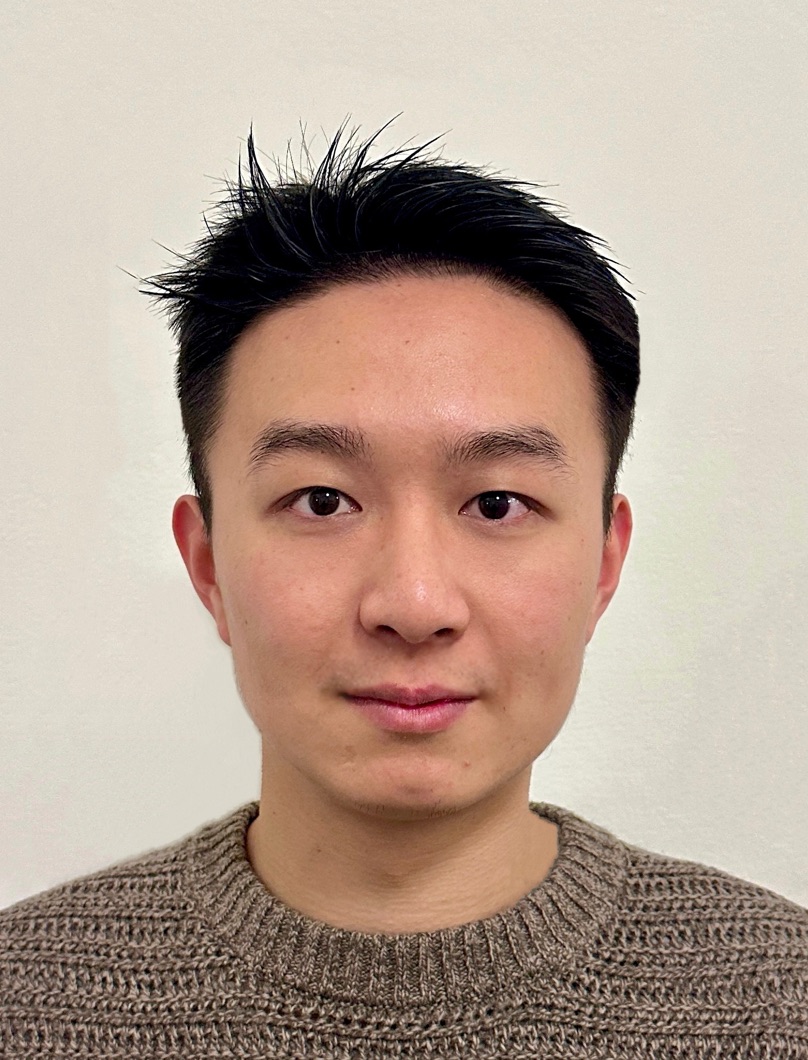}}]{Guang Chai}
	received the B.Sc. degree in Electrical Engineering from the University of Electronic Science and Technology of China (UESTC), and the M.Sc. degree in Electrical Engineering and Information Technology from the Technical University of Munich (TUM). Since April 2023, he has been pursuing the Ph.D. degree with the Technical University of Berlin (TUB), in collaboration with the Munich Research Center, Huawei Technologies Duesseldorf GmbH, Germany. His research interests include wireless communications and signal processing, with a focus on near-field sensing and codebook design for ISAC.
\end{IEEEbiography}
\vspace{-9mm}
\begin{IEEEbiography}[{\includegraphics[width=1in,height=1.25in,clip,keepaspectratio]{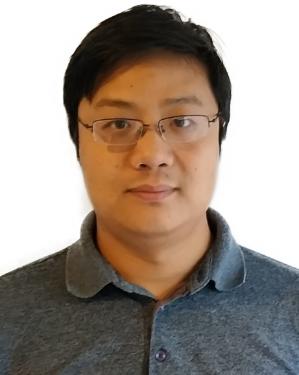}}]{Zhibin Yu}
	received the M.Sc. degree in Communication Engineering from the University of Ulm, Germany, in 2009, and a second M.Sc. degree in Embedded Systems from Eindhoven University of Technology, the Netherlands, in 2011. From 2011 to 2019, he worked as a Wireless System Engineer with Intel Deutschland GmbH. Since 2019, he has worked at Huawei Munich Research Center as a Senior Principal Research Engineer. He is the inventor or co-inventor of more than 100 patents. 
	%He received the Best Paper Award in IEEE Middle East Conference on Communications and Networking (MECOM), in 2024. 
	His current research interests are advanced signal processing methods for 5G/beyond 5G terminal modems, ISAC, and mmWave beamforming.
\end{IEEEbiography}
\vspace{-9mm}
\begin{IEEEbiography}[{\includegraphics[width=1in,height=1.25in,clip,keepaspectratio]{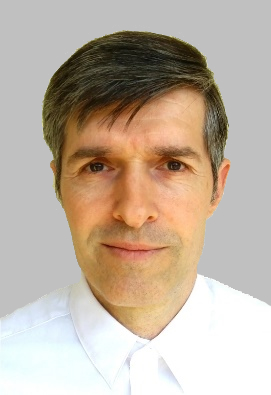}}]{Thomas Wagner}
		received the Dipl.-Ing. from Friedrich Alexander University Erlangen in 1996. He worked as a systems engineer at Philips Semiconductors, NXP Semiconductors, Ericsson and Intel before joining the Wireless Terminal Architecture and Design Department at Huawei's German research center in 2020.
	His current research interests include 5G modem architecture and wireless receiver algorithms including MIMO techniques.
	
\end{IEEEbiography}
\vspace{-9mm}
\begin{IEEEbiography}[{\includegraphics[width=1in,height=1.25in,clip,keepaspectratio]{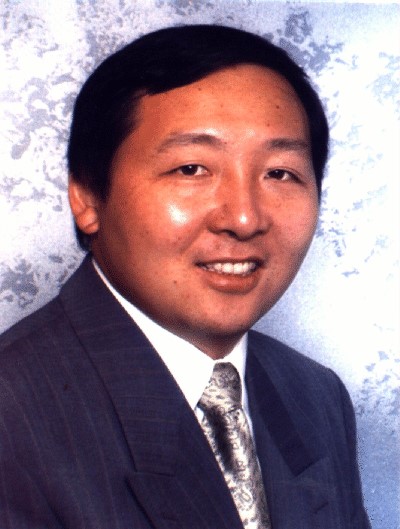}}]{Xiaofeng Wu}
		received the Dipl.-Ing. and Ph.D. degrees in electrical engineering from Ruhr University Bochum or the Technical University of Dortmund, Germany, in 1987 and 1993, respectively. He worked as a Wireless Principal Engineer and as the Engineering Manager with Infineon Technologies and Intel before joining Huawei’s German Research Center as the Director of the Wireless Terminal Chipset Technology Laboratory in 2019. 
	%He is with the Wireless Terminal Architecture and Design Department, HiSilicon. 
	He holds more than 80 filed patents. His current research interests include 5G/beyond 5G modem architecture, MIMO techniques, and mmWave communications.
\end{IEEEbiography}
\vspace{-9mm}
\begin{IEEEbiography}[{\includegraphics[width=1in,height=1.25in,clip,keepaspectratio]{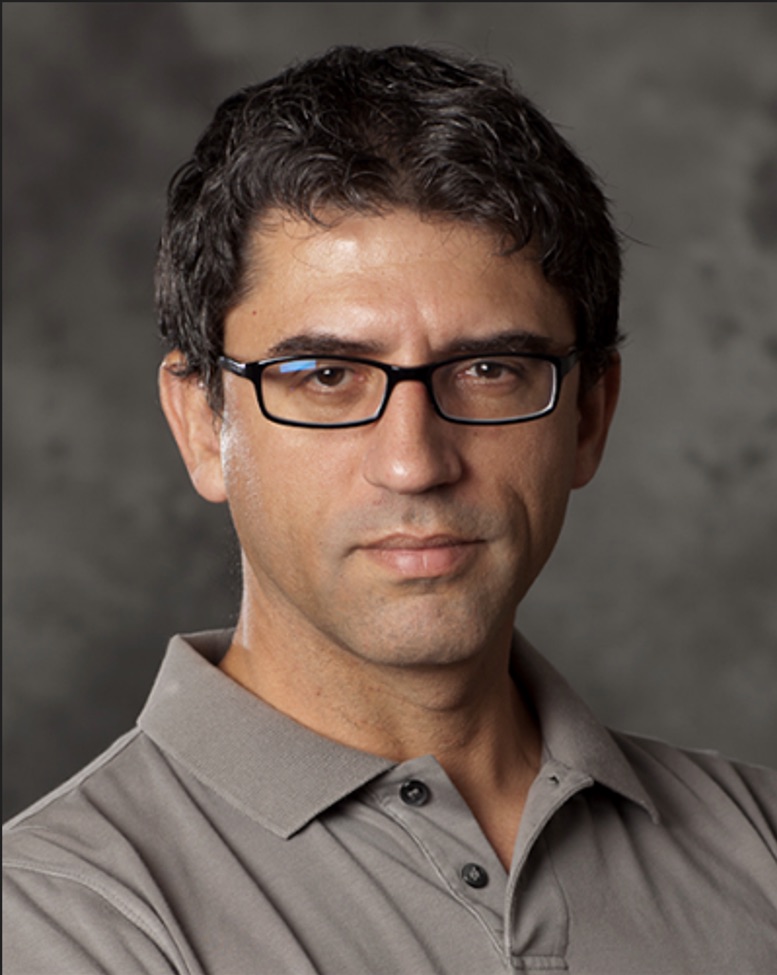}}]{Giuseppe Caire} 
   was born in Torino in 1965. He received a B.Sc. in Electrical Engineering from Politecnico di Torino in 1990, an M.Sc. from Princeton University in 1992, and a Ph.D. from Politecnico di Torino in 1994. He has been a postdoctoral research fellow with the European Space Agency (ESTEC, Noordwijk, The Netherlands) in 1994-1995, Assistant Professor in Telecommunications at the Politecnico di Torino, Associate Professor at the University of Parma, Italy, Professor with the Department of Mobile Communications at the Eurecom Institute, Sophia Antipolis, France, a Professor of Electrical Engineering with the Viterbi School of Engineering, University of Southern California, Los Angeles, and he is currently an Alexander von Humboldt Professor with the Faculty of Electrical Engineering and Computer Science at the Technical University of Berlin, Germany.
	%He received the Jack Neubauer Best System Paper Award from the IEEE Vehicular Technology Society in 2003, the IEEE Communications Society and Information Theory Society Joint Paper Award in 2004 and in 2011, the Okawa Research Award in 2006, the Alexander von Humboldt Professorship in 2014, the Vodafone Innovation Prize in 2015, an ERC Advanced Grant in 2018, the Leonard G. Abraham Prize for best IEEE JSAC paper in 2019, the IEEE Communications Society Edwin Howard Armstrong Achievement Award in 2020, the 2021 Leibniz Prize of the German National Science Foundation (DFG), and the CTTC Technical Achievement Award of the IEEE Communications Society in 2023. Giuseppe Caire is a Fellow of IEEE since 2005. He has served in the Board of Governors of the IEEE Information Theory Society from 2004 to 2007, and as officer from 2008 to 2013. He was President of the IEEE Information Theory Society in 2011. His main research interests are in the field of communications theory, information theory, channel and source coding with particular focus on wireless communications.
	He received the Alexander von Humboldt Professorship in 2014, an ERC Advanced Grant in 2018, the IEEE Communications Society Edwin Howard Armstrong Achievement Award in 2020, and the Leibniz Prize of the German National Science Foundation (DFG) in 2021. Giuseppe Caire is a Fellow of IEEE since 2005. He was President of the IEEE Information Theory Society in 2011. His main research interests are in the field of communications theory, information theory, channel and source coding with particular focus on wireless communications.
\end{IEEEbiography}

\vspace{2pt}

\vfill

\end{document}